\documentclass[12pt, a4paper]{article}

\usepackage{ulem}
\usepackage{amsfonts}
\usepackage{amsmath}
\usepackage{calc}
\usepackage{color}
\usepackage{epsfig}
\usepackage{float}
\usepackage{graphicx}
\usepackage{multirow}
\usepackage{natbib}
\usepackage{sidecap}
\usepackage{slashbox}
\usepackage{txfonts}
\usepackage{cancel}
\usepackage[normal]{caption}
\usepackage{nameref}
\usepackage{enumerate}
\usepackage{varioref}

\usepackage{sectsty}
\allsectionsfont{\normalsize}

\newcommand{\mbf}[1]{\mathbf{#1}}

\begin{document}
\normalsize
\bibliographystyle{elsart-harv-doi}

\setcitestyle{authoryear, round, semicolon, aysep={, }, yysep={, }, notesep={; }}

\parskip 15pt
\baselineskip 0.25in

\begin{center}
{\LARGE \bf Time and category information in pattern-based codes}
\end{center}
\vspace{-1.0cm}

\begin{center}
{\bf  Hugo Gabriel Eyherabide$^1$, and In\'es Samengo$^1$}
\end{center}
\vspace{-1.0cm}

\begin{enumerate}
\item[$^1$] Centro At\'omico Bariloche and Instituto Balseiro,
San Carlos de Bariloche, Argentina.
\end{enumerate}
\vspace{-0.5cm}

\noindent {\bf Correspondence}
\\Hugo Gabriel Eyherabide\\ Centro At\'omico Bariloche\\
San Carlos de Bariloche, (8400), R\'{\i}o Negro, Argentina.\\ Tel:
++ 54 2944 445100 (int: 5391 / 5345). Fax: ++54 2944 445299.\\
Email: eyherabh@ib.cnea.gov.ar

\noindent {\bf Running title:}\\ Time and category information.


\renewcommand{\thesection}{\arabic{section}.}
\renewcommand{\thesubsection}{\arabic{section}.\arabic{subsection}.}
\renewcommand{\thesubsubsection}{\arabic{section}.\arabic{subsection}.\arabic{subsubsection}.}

\labelformat{section}{section~\arabic{section}}
\labelformat{subsection}{section~\arabic{section}.\arabic{subsection}}
\labelformat{subsubsection}{section~\arabic{section}.\arabic{subsection}.\arabic{subsubsection}}

\renewcommand{\theenumi}{{\it\alph{enumi}}}
\renewcommand{\labelenumi}{{\it\alph{enumi}}-}

\parskip 15pt
\baselineskip 0.25in

\section*{Abstract}

Sensory stimuli are usually composed of different features (the \emph{what}) appearing at irregular
times (the \emph{when}). Neural responses often use spike patterns to represent sensory
information. The \emph{what} is hypothesised to be encoded in the identity of the elicited patterns
(the pattern categories), and the \emph{when}, in the time positions of patterns (the pattern
timing). However, this standard view is oversimplified. In the real world, the \emph{what} and the
\emph{when} might not be separable concepts, for instance, if they are correlated in the stimulus.
In addition, neuronal dynamics can condition the pattern timing to be correlated with the pattern
categories. Hence, timing and categories of patterns may not constitute independent channels of
information. In this paper, we assess the role of spike patterns in the neural code, irrespective
of the nature of the patterns. We first define information-theoretical quantities that allow us to
quantify the information encoded by different aspects of the neural response. We also introduce the
notion of synergy/redundancy between time positions and categories of patterns. We subsequently
establish the relation between the \emph{what} and the \emph{when} in the stimulus with the timing
and the categories of patterns. To that aim, we quantify the mutual information between different
aspects of the stimulus and different aspects of the response. This formal framework allows us to
determine the precise conditions under which the standard view holds, as well as the departures
from this simple case. Finally, we study the capability of different response aspects to represent
the \emph{what} and the \emph{when} in the neural response.

\section*{Keywords}

\noindent patterns, neural code, sensory encoding, information theory, redundancy, synergy,
stimulus features, feature extractor.

\pagebreak


\section{Introduction: Patterns in the neural response}\label{intro}

Sensory neurons represent external stimuli. In realistic conditions, different stimulus features
(for example, the presence of a predator or a prey) appear at irregular times. Therefore, an
efficient sensory system should not only represent the identity of each perceived stimulus, but
also, its timing. Colloquially, qualitative differences between stimulus features have been called
the \emph{what} in the stimulus, whereas the temporal locations of the features constitute the
\emph{when}. Spike trains can encode both the \emph{what} and the \emph{when}, for example, as a
sequence of spike patterns. This idea constitutes a standard view \citep{theunissen1995, borst1999,
krahe2004}, where the timing of patterns indicates \emph{when} stimulus features occur, while the
pattern identities tag \emph{what} stimulus features happened \citep{martinezconde2002, alitto2005,
oswald2007, eyherabide2008}. The information provided by the distinction between different spike
patterns is here called \emph{category information}. In the same manner, the information
transmitted by the timing of spike patterns is here called \emph{time information}. According to
the standard view, the category and the time information represent the knowledge of the \emph{what}
and the \emph{when} in the stimulus, respectively. In this work, we address the conditions under
which these assumptions hold, as well as departures from the standard view.

Many studies have shown the ubiquitous presence of patterns in the neural response. The patterns
can be, for instance, high-frequency burst-like discharges of varying length and latency. Examples
have been found in primary auditory cortex \citep{nelken2005}, the salamander retina
\citep{gollisch2008}, the mammalian early visual system \citep{debusk1997, martinezconde2002,
gaudry2008}, and grasshopper auditory receptors \citep{eyherabide2009, sabourin2009}. In other
cases, the patterns are spike doublets of different inter-spike interval (ISI) duration.
\citet{reich2000} presented an example of this type in primate V1; and \citet{oswald2007} found a
similar code in the electrosensory lobe of the weakly electric fish. In yet other cases, patterns
are more abstract spatiotemporal combinations of spikes and silences defined in single neurons
\citep{fellous2004} and neural populations \citep{nadasdy2000, gutig2006}.

If different spike patterns represent different stimulus features, which aspects of the pattern are
relevant to the distinction between the different features? To answer this question, previous
studies have classified the response patterns into different types of categories, depending on
different response aspects. The relevance of each candidate aspect was addressed using what we here
define as the category information. For example, in the auditory cortex, \citet{furukawa2002}
assessed how informative patterns were when categorised in three different ways, using the first
spike latency, the total number of spikes, or the variability in the spike timing. In an even more
ambitious study, \citet{gawne1996} have not only compared the information separately transmitted by
response latency and spike count, but also related these two response properties to two different
stimulus features: contrast and orientation, respectively. However, these works have not addressed
how the stimulus timing is represented by the response patterns.

The role of patterns in signaling the occurrence of the stimulus features can only be addressed in
those experiments where the stimulus features appear at irregular times. In this context, previous
approaches have estimated the time information \citep{gaudry2008, eyherabide2010}, or have either
employed other statistical measures such as reverse correlation \citep{martinezconde2000,
eyherabide2008}. The time information was calculated as the one encoded by the pattern onsets
alone, without distinguishing between different types of patterns.

In this paper, we analyse the role of timing and categories of patterns in the neural code. To this
aim, we build different representations of the neural response preserving one of these two aspects
at a time. This allows us to quantify the time and the category information separately. We
determine the precise meaning of these quantities and study of their variations for different
representations of the neural response. Unlike previous works \citep{gaudry2008, eyherabide2009,
foffani2009}, we quantify the information preserved and lost when the neural response is read out
in such a way that only the categories (timing) of patterns are preserved. As a result, the
relevance of each aspect of the neural response is unambiguously determined.

In principle, the timing and the categories of spike patterns may be correlated. These interactions
may be due to properties of the encoding neuron \citep[such as latency
codes][]{furukawa2002,gollisch2008}, properties of the decoding neuron \citep[when reading a
pattern-based code][]{lisman1997,reinagel1999}, the convention used to assigned a time reference to
the patterns \citep{nelken2005,eyherabide2008}, or the convention used to identify the patterns
from the neural response \citep{fellous2004, alitto2005, gaudry2008}. A statistical dependence
between timing and categories of patterns may, for example, introduce redundancy between the time
and category information. Thus, the same information may be contained in different aspects of the
response (categorical or temporal aspects). In addition, the statistical dependence might also
induce synergy, in which case extracting all the information about the \emph{what} and the
\emph{when} requires the simultaneous read-out of both aspects. The presence of synergy and
redundancy between the time and category information may affect the way each of them represents the
\emph{what} and the \emph{when} in the stimulus.

In the present study, we provide a formal framework to gain insight of the interaction between the
timing and the categories of patterns for different neural codes. We formally define the
\emph{what} and the \emph{when} as representations of the stimulus preserving only the identities
and timing of stimulus features, respectively. We then establish the conditions under which the
pattern categories encode the \emph{what} in the stimulus, and the timings the \emph{when}. We also
study departures from this standard interpretation, in particular, when the time position of
patterns depends on their internal structure. We show the impact of this dependence on both the
link with the \emph{what} and the \emph{when} and the relative relevance of the timing and
categories of patterns. Our study is therefore intended to motivate more systematic explorations of
the neural code in sensory systems.

\section{Methods}\label{methods}

\subsection{Reduced representations of the neural response}\label{methods:representations}

\begin{figure}[htp]
\includegraphics{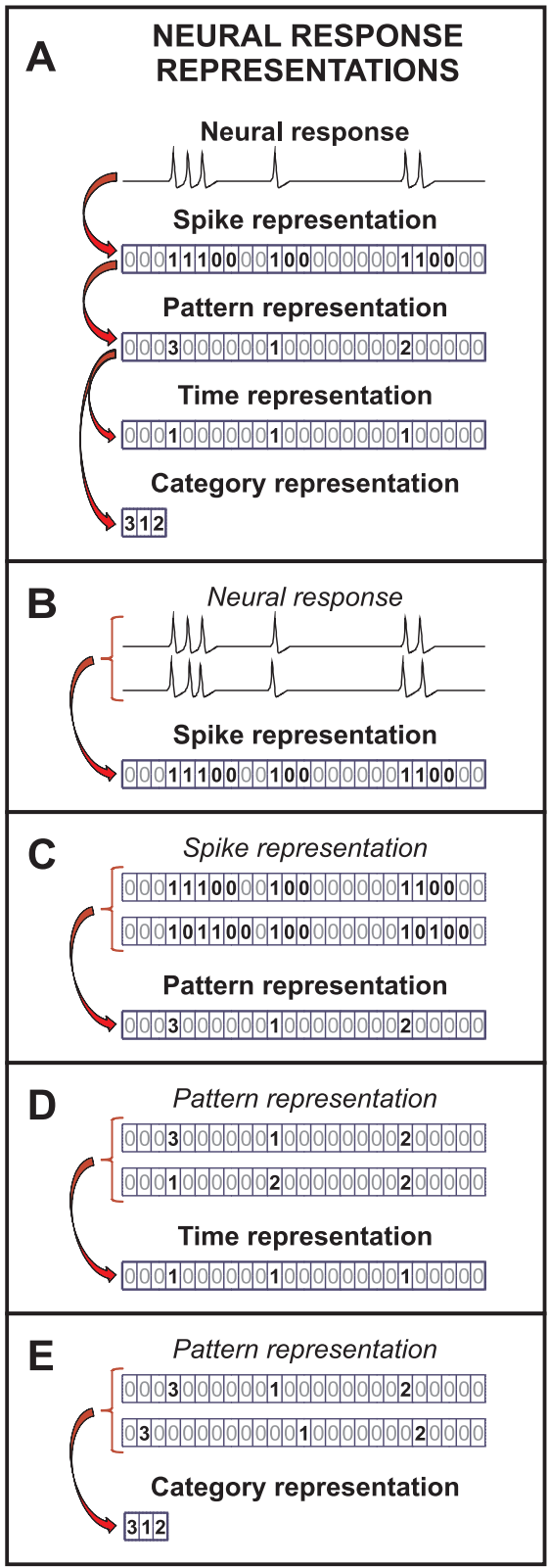}
\caption{\label{f1} {\bf Representations of the neural response.} ({\bf A}) In the spike
representation, only the timing of action potentials is described, discarding the fine structure of
the voltage traces. In the pattern representation, only the timing and categories of spike patterns
remain. This representation is further transformed, to obtain the time and the category
representations. The time (category) representation only keeps information about the timing
(categories) of the spike patterns. ({\bf B}), ({\bf C}), ({\bf D}) and ({\bf E}) Each successive
transformation of the neural response through a deterministic function simultaneously reduces both
the variability in the neural response and number of possible responses.}
\end{figure}

A \emph{representation} is a description of the neural response. Formally, it is obtained by
transforming the recorded neural activity through a deterministic mapping. Throughout this paper,
the expressions ``deterministic mapping" and ``function" are used as synonyms. We only consider
functions that transform the unprocessed neural response $\mbf{U}$ into sequences of events
$e_i=(t_i, c_i)$, characterised by their time positions ($t_i$) and categories ($c_i$). An event is
a definite response stretch. Based on their internal structure, events are classified into
different categories, as explained later in this section. Individual spikes may be regarded as the
simplest events. In this case, the sequence of events is called the \emph{ spike representation}
(see Figure~\ref{f1}A), comprising events belonging to a single category: the category ``spikes".

From the spike representation, we can define more complex events, hereafter called \emph{patterns}
(see bold symbols in the spike representation in Figure~\ref{f1}A). Patterns may be defined in
terms of spikes, bursts or ISIs \citep{alitto2005, luna2005, oswald2007, eyherabide2008}. They may
involve one or several neurons. Examples of population patterns are coincident firing, precise
firing events and sequences, or distributed patterns \citep{hopfield1995, abeles2001, reinagel2002,
gutig2006}. The sequence of patterns obtained by transforming the spike representation is called
the \emph{pattern representation}. Analogously, the sequence of patterns only characterised by
either their time positions or their categories constitute the \emph{time representation} and
\emph{category representation}, respectively. Details on how to build these sequences are explained
below. For simplicity, these sequences are represented in Figure~\ref{f1} as sequences of symbols
$n$, indicating specific events ($n>0$) and silences ($n=0$).

Formally, to obtain the spike representation ($\mbf{R}$), the unprocessed neural response
($\mbf{U}$) is transformed into a sequence of spikes (1) and silences (0) (Figure~\ref{f1}A). The
time bin is taken small enough to include at most one spike. Differences in shape of action
potentials are ignored, while their time positions are preserved, with temporal precision limited
by the bin size. As a result, several sequences of action potentials may be represented by the same
spike sequence (see Figure~\ref{f1}B).

In the pattern representation ($\mbf{B}$), the spike sequence is transformed into a sequence of
silences ($n=0$) and spike patterns ($n=b>0$), distinguished solely by their category $b$. For
example, in Figure~\ref{f1}, patterns are defined as response stretches containing consecutive
spikes separated by at most one silence. The time positions of the pattern is defined as the first
spike in each pattern stretch, whereas patterns with the same number of spikes are grouped into the
same pattern category. Only information about pattern categories and time positions remains
(compare the bold symbols in the spike and the pattern representation in Figure~\ref{f1}.A). By
ignoring differences among patterns within categories, several spike sequences can be mapped into
the same pattern sequence, as shown in Figure~\ref{f1}C.

The time position of patterns is measured with respect to a common origin, in general, the
beginning of the experiment. It can be defined, for example, as the first (or any other) spike of
the pattern or as the mean response time \citep{lisman1997, nelken2005, eyherabide2009}. Patterns
are classified into categories according to different aspects describing their internal structure,
such as the latency, the number of spikes or the spike-time dispersion \citep{gawne1996,
theunissen1995, furukawa2002}. Notice that latencies are usually defined with respect to the
stimulus onset, which is not a response property \citep{chase2007, gollisch2008}. Thus, latencies
and timing of spike patterns are different concepts, and the latency cannot be read out from the
neural response alone. However, latencies have also been defined with respect to the local field
potential \citep{montemurro2008} or population activity \citep{chase2007}. These definitions can be
regarded as internal aspects of spatiotemporal spike patterns \citep{theunissen1995, nadasdy2000}.

Categories of patterns can be built by discretizing the range of one or several internal aspects.
For example, \citet{reich2000} defined patterns as individual ISIs, and categorised them in terms
of their duration. Three categories were considered, depending on whether the ISI was short, medium
or large. In other cases, patterns may be sequences of spikes separated by less than a certain time
interval. Categories of patterns can then be defined, depending on the number of spikes in each
pattern \citep{reinagel2000, martinezconde2002, eyherabide2010}, as shown in Figure~\ref{f1}, or
depending on the length of the first ISI \citep{oswald2007}. The theory developed in this paper is
valid irrespective of the way in which one chooses to define the pattern time positions and the
pattern categories.

From the pattern sequence, we obtain the time representation ($\mbf{T}$) by only keeping the time
positions of patterns. As a result, the neural response is transformed into a sequence of silences
(0) and events (1), indicating the occurrence of a pattern in the corresponding time bin and
disregarding its category. The temporal precision of the pattern representation is preserved in the
time representation. However, by ignoring differences between categories, different pattern
sequences can be mapped into the same time representation, as illustrated in Figure~\ref{f1}D.

The category representation ($\mbf{C}$) is complementary to the time representation. It is obtained
from the pattern sequence, by only keeping information about the categories of patterns while
ignoring their time positions. The neural response is transformed into a sequence of integer
symbols $n>0$, representing the sequence of pattern categories in the response. The exact time
position of patterns is lost: only their order remains. Therefore, several pattern sequences may be
mapped onto the same category sequence, as indicated in Figure~\ref{f1}E.

The spike ($\mbf{R}$), pattern ($\mbf{B}$), time ($\mbf{T}$) and category ($\mbf{C}$)
representations are derived through functions that depend only on the previous representation, as
denoted by the arrows in Figure~\ref{f1}A, and formally expressed by the following equations:

\begin{subequations}\label{representations}
\begin{align}
&\text{Neural response}  &{}&\mbf{U}\, \text{(experiment)}\\
&\text{Spike representation}  &{}&\mbf{R}=h_{\mbf{U}\rightarrow\mbf{R}}\left(\mbf{U}\right)\label{hUR}\\
&\text{Pattern representation}  &{}&\mbf{B}=h_{\mbf{R}\rightarrow\mbf{B}}\left(\mbf{R}\right)\label{hRB}\\
&\text{Time representation}  &{}&\mbf{T}=h_{\mbf{B}\rightarrow\mbf{T}}\left(\mbf{B}\right)\label{hBT}\\
&\text{Category representation}
&{}&\mbf{C}=h_{\mbf{B}\rightarrow\mbf{C}}\left(\mbf{B}\right)\label{hBC}\, ;
\end{align}
\end{subequations}

\noindent where $h_{\mbf{X}\rightarrow\mbf{Y}}$ represents the function $h$ that is applied to the
representation $\mbf{X}$ to obtain the representation {\bf Y}. These transformations progressively
reduce both the variability in the neural response and the number of possible responses

\begin{subequations}
\begin{equation}
H(\mbf{U})\geq H(\mbf{R})\geq H(\mbf{B})\geq \left\{
\begin{array}{l}
H(\mbf{T})\\
H(\mbf{C})
\end{array}\right. \, ;\\ \label{entropyreprel}
\end{equation}
\begin{equation}
|\mbf{U}|\geq |\mbf{R}|\geq |\mbf{B}|\geq \left\{
\begin{array}{l}
|\mbf{T}|\\
|\mbf{C}|
\end{array}\right. \, ;
\end{equation}
\end{subequations}

\noindent where $H(\mbf{X})$ means the entropy $H$ of the set $\mbf{X}$ \citep{coverthomas}, and
$|\mbf{X}|$ indicates its cardinality, i.e. the number of elements of the set $\mbf{X}$.

\subsection{Calculation of mutual information rates} \label{infocalculo}

The \emph{mutual information} $\mathrm{I}(\mbf{X};\mbf{S})$ between two random variables $\mbf{X}$
and $\mbf{S}$ is defined as the reduction in the uncertainty of one of the random variables due to
the knowledge of the other. It is formally expressed as a difference between two entropies

\begin{equation}\label{infodef}
\mathrm{I}(\mbf{X};\mbf{S}) = \mathrm{H}(\mbf{X})-\mathrm{H}(\mbf{X}|\mbf{S})\, ;
\end{equation}

\noindent where $\mathrm{H}(\mbf{X})$ is the \emph{total entropy} of $\mbf{X}$ and
$\mathrm{H}(\mbf{X}|\mbf{S})$ represents the \emph{conditional} or \emph{noise entropy} of
$\mbf{X}$ provided that $\mbf{S}$ is known \citep{coverthomas}.

We estimate the mutual information between the stimulus $\mbf{S}$ and a representation $\mbf{X}$ of
the neural response using the so-called \emph{Direct Method}, introduced by \citet{strong1998}. The
unprocessed neural response $\mbf{U}$ is divided into time intervals $\mbf{U}_{\tau}$ of length
$\tau$. Each response stretch $\mbf{U}_{\tau}$ is then transformed into the discrete-time
representation $\mbf{X}_{\tau}$
($\mbf{X}_{\tau}=h_{\mbf{U}\rightarrow\mbf{X}}\left(\mbf{U}_{\tau}\right)$), also called
\emph{words}. As a result

\begin{equation}
I(\mbf{S};\mbf{V}_{\tau})\geq I(\mbf{S};\mbf{X}_{\tau})\, .
\end{equation}

\noindent This inequality is valid for every time interval of length $\tau$ \citep{coverthomas} and
is not limited to the asymptotic regime for long time intervals, like in previous calculations
\citep{gaudry2008, eyherabide2009}. The mutual information calculated with words of length $\tau$
only quantifies properly the contribution of spike patterns that are shorter than $\tau$. In order
to include the correlations between these patterns, even longer words are needed. Therefore, in
this study, the maximum window length ranged between 3 and 4 times the maximum pattern duration.

The total entropy ($\mathrm{H}(\mbf{X}_{\tau})$) and noise entropy
($\mathrm{H}(\mbf{X}_{\tau}|\mbf{S})$) are estimated using the distributions of words
$\mbf{X}_{\tau}$ unconditional ($P(\mbf{X}_{\tau})$) and conditional ($P(\mbf{X}_{\tau}|\mbf{S})$)
on the stimulus $\mbf{S}$, respectively. The mutual information
$\mathrm{I}(\mbf{S};\mbf{X}_{\tau})$ is computed by subtracting
$\mathrm{H}(\mbf{X}_{\tau}|\mbf{S})$ from $\mathrm{H}(\mbf{X}_{\tau})$ (Eq.~\ref{infodef}). This
calculation is repeated for increasing word lengths, and the \emph{mutual information rate}
$I(\mbf{S};\mbf{X})$ between the stimulus $\mbf{S}$ and a representation $\mbf{X}$ of the neural
response is estimated as

\begin{equation}\label{mutualinfo}
I(\mbf{S};\mbf{X})=\lim_{{\tau}\rightarrow\infty}{\frac{\mathrm{I}(\mbf{S};\mbf{X}_{\tau})}{{\tau}}}\,
.
\end{equation}

\noindent This quantity represents the mutual information per unit time when the stimulus and the
response are read out with very long words. In this work we always calculate mutual information
\emph{rates} unless it is otherwise indicated. However, for compactness, we sometimes refer to this
quantity simply as ``information''.

The estimation of information suffers from both bias and variance \citep{panzeri2007}. In this
work, the sampling bias of the information estimation was corrected using the NSB approach for the
experimental data \citep{nemenman2004}. For the simulations, we used instead the quadratic
extrapolation \citep{strong1998}, due to its simplicity and the possibility of generating large
amounts of data. The standard deviation of the information was estimated from the linear
extrapolation to infinitely long words \citep{rice1989}. The bias correction was always lower than
$1.5\, \%$ and the standard deviation, always lower than $1\, \%$, for all simulations and all word
lengths; thus error bars are not visible in the figures. When comparisons between information
estimations were needed, one-sided t-tests were performed \citep{rice1989}.

\subsection{Simulated data}\label{methods:simdata}

Simulations are used to exemplify the theoretical results and to gain additional insight on how
different response conditions affect information transmission in well-known neural models and
neural codes. They represent highly idealised cases, with unrealistically long runs and number of
trials, that allow us to readily exemplify the theoretical results and transparently obtain
reliable information estimates. Firstly, we define the parameters used in the simulations and
relate them to the specific aspects of the stimulus and the response. Then, we report the specific
values for the parameters.

\subsection{General description}

In the simulations, the stimulus consists of a random sequence of instantaneous discrete events,
here called \emph{stimulus features}. Each stimulus feature is characterised by specific physical
properties, as for example, the colour of a visual stimulus, the pitch of an auditory stimulus, the
intensity of a tactile stimulus, or the odour of an olfactory stimulus \citep{poulos1984,
rolen2007, nelken2008, mancuso2009}. In the real world, however, features are not necessarily
discrete. If they are continuous, one can discretize them by dividing their domain into discrete
categories \citep{martinezconde2002, eyherabide2008, marsat2009}. The present framework sets no
upper limit to the number of features, nor to the similarity between different categories. In
addition, features might not be instantaneous but rather develop in extended time windows, as it
happens with the chirps in the weakly electric fish \citep{benda2005}, the oscillations in the
electric field potential \citep{oswald2007} and the amplitude of auditory stimuli
\citep{eyherabide2008}. In order to capture the duration of real stimuli, in the simulations we
define a \emph{minimum inter-feature interval} $\lambda^s_{min}$, for each feature $s$. After the
presentation of a feature $s$, no other feature may appear in an interval lower or equal to
$\lambda^s_{min}$.

\begin{figure}[htp]
\includegraphics{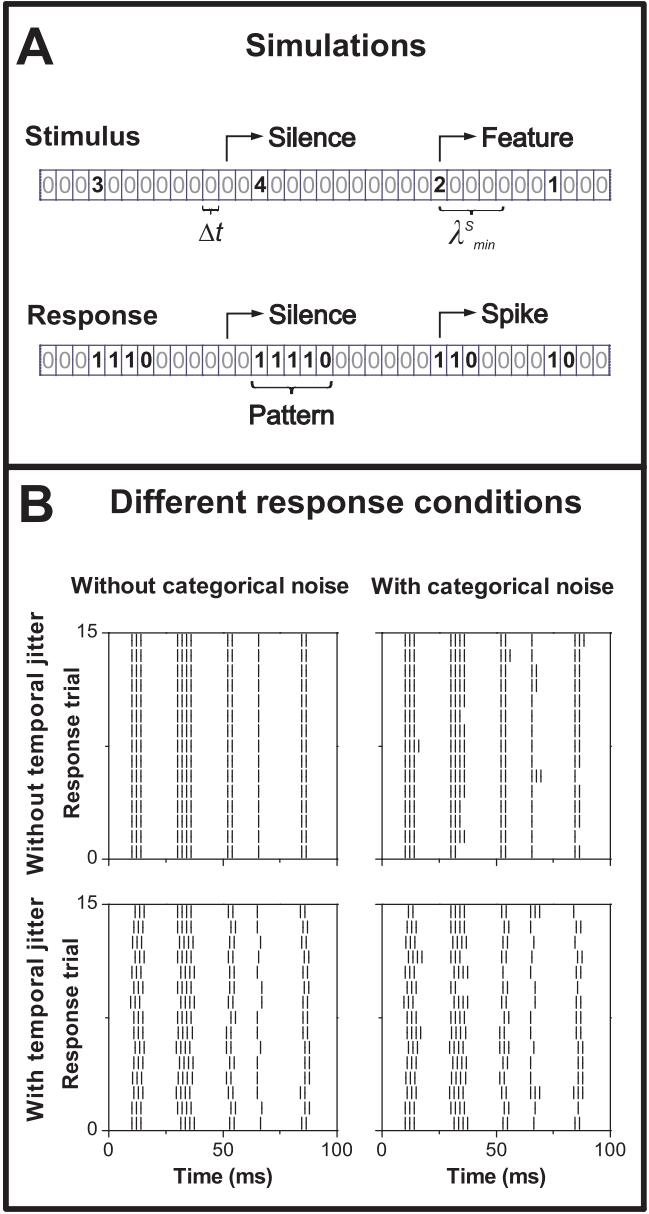}
\caption{\label{f2} {\bf Simulations: Design and construction.} ({\bf A}) Example of a stimulus
stretch and the elicited response. The stimulus is depicted as an integer sequence of silences
($0$) and features ($s>0$), one symbol per time bin of size $\Delta t$. After a feature arrival,
the stimulus remains silent for a period $\lambda^S_{min}$. The response is represented as a binary
sequence of spikes ($1$) and silences ($0$). Each stimulus feature elicits a response pattern: A
burst containing $n$ spikes. Different categories correspond to different intra-burst spike counts.
({\bf B}) Examples of different response conditions. \emph{Upper panels}: no temporal jitter;
\emph{lower panels}: the pattern, as a whole, is displaced due to temporal jitter; \emph{left
panels}: no categorical noise; \emph{right panels}: each stimulus feature elicits pattern responses
belonging to more than a single category.}
\end{figure}

In the simulated data, each stimulus feature elicits a neural response (see Figure~\ref{f2}A).
Since in this paper we are interested in pattern-based codes, each feature generates a pattern of
spikes belonging to some pattern category. The correspondence between stimulus features and pattern
categories may be noisy. We consider both categorical noise (the pattern category varies from trial
to trial) and temporal noise (the timing of the pattern varies from trial to trial). In
Figure~\ref{f2}B, we show examples of all noise conditions using burst-like response patterns. In
those examples, categories were defined according to the number of spikes in each burst.

Symbolically, the stimulus $\mbf{S}$ is represented as a sequence of symbols $s$, one per time bin
$\Delta t$. Each $s$ is drawn randomly from the set of all possible outcomes $\Sigma_{\mbf{s}}=\{0,
1, \ldots, N_S\}$. The symbol $s = 0$ indicates a silence (the absence of a feature), whereas $s >
0$ tags the presence of a given feature. Each feature $s$ elicits a response pattern $\mbf{r}$,
drawn from the set $\Sigma_{\mbf{r}}$ of all possible patterns, with probability
$P_{\mbf{r}}(\mbf{r}|s)$. The response pattern $\mbf{r}$ may appear with latency $\mu_{\mbf{r}}$,
which might depend on the evoked pattern $\mbf{r}$. A neural response $\mbf{R}$, elicited by a
sequence of stimulus features, may be composed of several response patterns (see bold symbol
sequences in \ref{f2}A).

Figure~\ref{f2}B shows example neural codes with no noise (upper left panel), categorical noise
alone (upper right), temporal noise alone (lower left), and a mixture of categorical and temporal
noise (lower right). The categorical noise is defined by $P_{\mbf{b}}(b|s)$, quantifying the
probability that a response category $b$ be elicited in response to stimulus $s$ (see \ref{appxA}
for the relation between $P_{\mbf{b}}(b|s)$ and $P_{\mbf{r}}(\mbf{r}|s)$). The temporal noise is
implemented as jitter in the pattern onset time. That is, temporal jitter affects the pattern as a
whole, displacing all spikes in the pattern by the same amount of time. The temporal displacement
is drawn from a uniform distribution in the interval $(-\sigma_b, \sigma_b)$, where the jitter
$\sigma_b$ may depend on the pattern $b$.

\subsubsection{Details and parameters}\label{detailsim}

Simulated neural responses consisted of four different patterns, elicited by a stimulus with four
different features. The response patterns were bursts of spikes, containing between 1 to 4 spikes.
The intra-burst ISI was $\gamma_{min}=2\, ms$. However, since the neural response is transformed
into the pattern representation, the results are valid irrespective of the nature of the patterns
(see \ref{methods:representations}). The stimulus was presented $200$ times, each one lasting for
$2000\, s$. The minimum inter-feature time interval is $\lambda_{min}=12\, ms$. In all cases, no
interference between patterns was considered (see \ref{departures}). We used a time bin of size
$\Delta t=1\, ms$.

\paragraph{Simulation 1:}
This simulation is used to illustrate the effect of using different representations of the neural
response, and to compare an ideal situation where the correspondence between features and patterns
is known, with a more realistic case, where the neural code is unknown. The temporal jitter was
$\sigma=1\, ms$ and the latency was $\mu=1\, ms$. Stimulus features probability $p(s)$ were set to:
$p(1)=0.06$, $p(2)=0.04$, $p(3)=0.03$, $p(4)=0.02$. Categorical noise ($p(b|s)$, $b\neq s$):
$p(i+1|i)=0.1\, (4-i)$, $0<i<4$; otherwise $p(b|s)=0$.

\paragraph{Simulation 2:}
These simulations are used to address the role of the timing and category of patterns in the neural
code, and to study the relation with the \emph{what} and the \emph{when} in the stimulus. The
latency was $\mu=1\, ms$. When present, temporal jitter was set to  $\sigma=1\, ms$ and categorical
noise ($p(b|s)$, $b \neq s$) was given by: $p(i+1|i)=p(i|i+1)=p(3|1)=p(2|4)=0.1$, $0<i<4$;
otherwise $p(b|s)=0$. Stimulus features probability $p(s)=0.025$, $0<s\leq 4$.

\subsection{Electrophysiology}\label{expdata}

\begin{figure}[htp]
\includegraphics{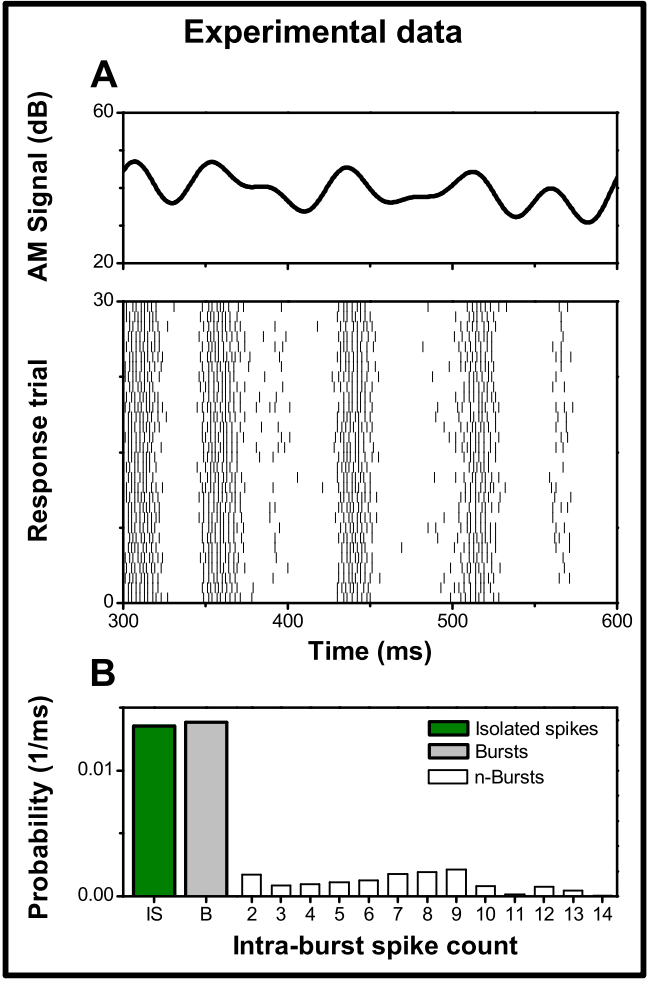}
\caption{\label{f3} {\bf Experimental data from a grasshopper auditory receptor neuron.} ({\bf A})
\emph{Upper panel:} Sample of the amplitude modulation of the sound stimulus used in the
recordings. \emph{Lower panel:} Response to 30 of 479 repeated stimulus presentations showing
conspicuous burst activity. Each vertical line represents a single spike. ({\bf B}) Probability of
firing a burst with $n$ intra-burst spikes, in a time bin of size $\Delta t=1\, ms$. Isolated
spikes ($n=1$) and burst activity ($n>1$) represent $49.4\%$ and $50.6 \%$ of the firing events,
respectively.}
\end{figure}

Experimental neural data were provided by Ariel Rokem and Andreas V. M. Herz; they performed
intracellular recordings \emph{in vivo}, on the auditory nerve of \emph{Locusta Migratoria}
\citep[see][for details]{rokem2006}. Auditory stimuli consisted of a $3\, kHz$ carrier sine wave,
amplitude modulated by a low pass filtered signal with a Gaussian distribution. The AM signal had a
mean amplitude of $53.9\, dB$, a $6\, dB$ standard deviation and a cut-off frequency of $25\, Hz$
(see Figure \ref{f3}A upper cell). Each stimulation lasted for $1000\, ms$ with a pause of $700\,
ms$ between repeated presentations of the stimulus, in order to minimise the influence of slow
adaptation. To eliminate fast adaptation effects, the first $200\, ms$ of each trial were
discarded. The recorded response (see Figure \ref{f3}A lower panel) consisted of 479 trials, with a
mean firing rate of $108 \pm 6\, ^{spikes}/_s$ (mean $\pm$ standard deviation across trials). Burst
activity was observed and associated with specific features in the stimulus \citep[see][for the
analysis of burst activity in the whole data set]{eyherabide2008}. Bursts contained up to 14
spikes; Figure \ref{f3}B shows the firing probability distribution as a function of the intra-burst
spike count.

\section{Results}

\subsection{Information transmitted by different representations of the neural response: spike and pattern information}\label{SpikePatternInfo}

In order to understand how stimuli are encoded in the neural response, the recorded neural activity
$\mbf{U}$ is transformed into several different representations. Each representation keeps some
aspects of the original neural response while discarding others. The spike representation $\mbf{R}$
is probably the most widely used (see \ref{methods:representations}). We define the \emph{spike
information} $I(\mbf{S};\mbf{R})$ as the mutual information rate between the stimulus $\mbf{S}$ and
the spike representation $\mbf{R}$ of the neural response.

The spike sequence can be further transformed into a sequence of patterns of spikes, called the
pattern representation $\mbf{B}$. To that end, all possible patterns of spikes are classified into
pre-defined categories, for example, burst codes, ISI codes, etc. (see
\ref{methods:representations} and references therein). We define \emph{pattern information}
$I(\mbf{S};\mbf{B})$ as the information about the stimulus $\mbf{S}$, carried by the sequence of
patterns $\mbf{B}$.

The pattern information cannot be greater than the spike information, which in turn cannot be
greater than the information in the unprocessed neural response

\begin{equation}\label{dpisp}
I(\mbf{S};\mbf{B}) \leq I(\mbf{S};\mbf{R}) \leq I(\mbf{S};\mbf{U}) \, .
\end{equation}

\noindent This result can be directly proved from the deterministic relation between $\mbf{U}$,
$\mbf{R}$ and $\mbf{B}$ (Eqs. \ref{representations}) and the \emph{data processing inequality}
\citep{coverthomas}. Notwithstanding, several neuroscience papers have reported data contradicting
Eq.~\ref{dpisp} (see \ref{dpidiscuss}). Intuitively, out of all the information carried by the
unprocessed neural response, the spike information only contains the information preserved in the
spike timing. Analogously, out of the information carried in the spike representation, the pattern
information only preserves the information carried by both the time positions and the categories of
the chosen patterns.

\subsection{Choosing the pattern representation}

In this paper, we quantify the amount of time and category information encoded by pattern-based codes. This information depends critically on the choice of the pattern representation. In this subsection, we discuss how to evaluate whether a given choice is convenient or not. One can choose any set of pattern categories to define the alphabet of the pattern
representation. Some choices, however, preserve more information about the stimulus than others. The comparison between the information carried by different pattern representations gives insight on how relevant to information transmission the preserved structures are \citep{victor2002, nelken2007}, i.e. formally, on whether they constitute
sufficient statistics \citep{coverthomas}. A suitable representation should reduce the variability in the neural response due to noise, while preserving the variability associated with variations in the encoded stimulus. Thus, any representation preserving less information than the spike information is neglecting informative variability. In addition, one may also be
interested in a neural representation that can be easily or rapidly read out, or that is robust to environmental changes, etc. The chosen neural representation typically results from a trade-off between these requirements.

\begin{figure}[htbp!]
\centering\includegraphics{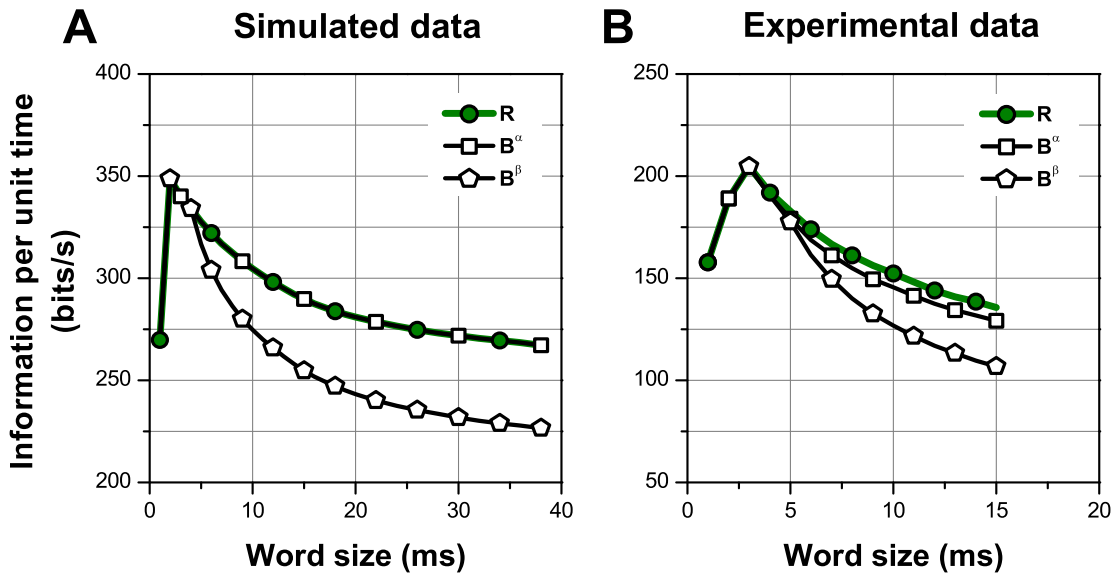} \caption{\label{f4}{\bf Information per unit
time transmitted by different choices of patterns.} The spike representation ($\mbf{R}$) is
transformed into a sequence of patterns grouped in categories according to the intra-pattern spike
count ($\mbf{B}^{\alpha}$), which is further transformed into a sequence of patterns classified as
isolated spikes or complex patterns ($\mbf{B}^{\beta}$). Comparing the amount of information
transmitted gives insight about the relevance of the structures preserved in the representations.
({\bf A}) Simulation of a neural response with four different patterns, elicited by a stimulus with
four different features, in presence of temporal jitter and categorical noise (see \ref{detailsim}
Simulation 1 for details). ({\bf B}) Experimental data from a grasshopper auditory receptor neuron.
In all cases, error bars $< 1\%$ (smaller than the size of the data points).}
\end{figure}

Here we focus on analysing whether the chosen representation alters the correspondence between the stimulus and the response. For us, a good representation is one where the informative variability is preserved, and the non-informative variability is discarded. As an example, we analyse two different situations (Figure~\ref{f4}). In panel \emph{A}, we use simulated data, where we know exactly how the neural code is structured. We can therefore compare the performance of the spike representation, with two pattern representations: one of them intentionally tailored to capture the true neural code that generated the data, and another representation discarding some informative variability. The neural response consists of a sequence of four different patterns, associated with each of four stimulus features, in the presence of temporal jitter and categorical noise (see \ref{detailsim} Simulation 1). In panel \emph{B}, we study experimental data (see \ref{expdata}), so the neural code is unknown.  Therefore, in this case we compare the spike representation with two candidate pattern representations, ignoring a-priori which is the most suitable.

For both simulation and experimental data, we estimated the information conveyed by the spike representation $\mbf{R}$; a pattern representation $\mbf{B}^{\alpha}$, where all bursts are grouped into categories according to their intra-burst spike count; and a second pattern representation $\mbf{B}^{\beta}$, with only two categories comprising isolated spikes and complex patterns. This is shown in Figure~\ref{f4}, where the information per unit time is plotted as a
function of the window size used to read the neural response. The representations are related
through functions, in such a way that $\mbf{B}^{\beta}$ is a transformation of $\mbf{B}^{\alpha}$,
which is in turn a transformation of $\mbf{R}$. Therefore, $I(\mbf{S};\mbf{B}^{\beta}) \leq
I(\mbf{S};\mbf{B}^{\alpha}) \leq I(\mbf{S};\mbf{R})$, for all finite response windows (see Eq.
\ref{dpisp}). Nevertheless, notice that $\mbf{B}^{\beta}$ may be a faster-to-read code than $\mbf{B}^{\alpha}$, since the latter requires a time window long enough to distinguish not only the differences between isolated spikes and bursts, but also the differences among bursts of different categories.

In the simulation (Figure~\ref{f4}A), the information carried by $\mbf{B}^{\alpha}$ is equal to the
spike information ($I_{Sim}(\mbf{S};\mbf{R})=I_{Sim}(\mbf{S};\mbf{B}^{\alpha})=254.2\pm 0.2\,
bits/s$, one-sided t-test, $p(10) = 0.5$). This is expected since, by construction, the neural code
used in the simulations is, indeed, $\mbf{B}^{\alpha}$. Therefore, in this case, $\mbf{B}^{\alpha}$
is a lossless representation. The choice of an adequate representation is more difficult in the
experimental example (Figure \ref{f4}B), where the neural code is not known beforehand. In this
case, $\mbf{B}^{\alpha}$ preserves less information than the spike sequence
($I_{Exp}(\mbf{S};\mbf{R})=133\pm 4 \, bits/s$, $I_{Exp}(\mbf{S};\mbf{B}^{\alpha})=121\pm 3 \,
bits/s$, one-sided t-test, $p(10)=0.004$). The information $I(\mbf{S};\mbf{B}^{\alpha})$ represents
$91\, \%$ of the spike information. In general, whether this amount of information is acceptable or
not depends on whether the loss is compensated by the advantages of attaining a reduced
representation of the response \citep{nelken2007}.

Distinguishing only between isolated spikes and bursts ($\mbf{B}^{\beta}$) diminishes the
information considerably in both examples (one-sided t-test, $p(10)<0.001$, both cases). In the
simulation, the information carried by $\mbf{B}^{\beta}$ is
$I_{Sim}(\mbf{S};\mbf{B}^{\beta})=208.7\pm 0.6 \, bits/s$, representing about $82.1 \, \%$ of the
spike information. This is expected since, by construction, different stimulus features are encoded
by different patterns. For the experimental data, $I_{Exp}(\mbf{S};\mbf{B}^{\beta})=91\pm 7 \,
bits/s$, representing about $68\, \%$ of the spike information. In both examples, the
representation $\mbf{B}^{\alpha}$ is ``more sufficient" than $\mbf{B}^{\beta}$. The difference
$I(\mbf{S};\mbf{B}^{\alpha})-I(\mbf{S};\mbf{B}^{\beta})$ constitutes a quantitative measure of the
role of distinguishing between bursts of $2,\,3,\,\ldots\, ,\,n$ spikes, provided that the
distinction between isolated spikes and bursts has already been made
($I(\mbf{S};\mbf{B}^{\alpha}|\mbf{B}^{\beta})$). However, $\mbf{B}^{\alpha}$ still preserves other
response aspects, such as pattern timing, number of patterns, etc. In what follows, we study the
role of different response aspects in information transmission.

\subsection{Informative aspects of the neural response}\label{results:infoaspects}

\begin{figure}[ht!]
\centering
\includegraphics{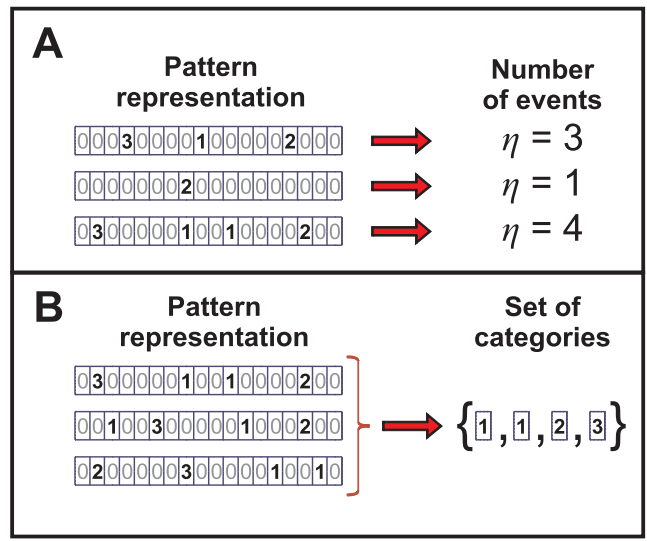}
\caption{\label{f5} {\bf Identifying the information carriers in the neural response.} ({\bf A})
The representation $\eta$ of the neural response is obtained by transforming the pattern sequence
such that only the number of events is preserved. ({\bf B}) By transforming the pattern sequence
into the representation $\Theta$, the information about the categories present in the neural
response is preserved, while their order of occurrence is disregarded. }
\end{figure}

The pattern representation may preserve one or several aspects of the neural response that could,
in principle, encode information about the stimulus. More specifically, if the response is analysed
using windows of duration $\tau$, there are several candidate response aspects that might be
informative, namely:

\begin{enumerate}
\item the number of patterns in the window (number of events - Figure~\ref{f5}A)\label{aspnoe}
\item the precise timing of each pattern in the window (time representation - Figure~\ref{f1}D)\label{asptr}
\item the pattern categories present in the window with no specification of their ordering (response set of categories - Figure~\ref{f5}B)\label{aspsoc}
\item the temporally-ordered pattern categories in the window (category representation - Figure~\ref{f1}E). \label{aspcr}
\end{enumerate}

We find that these aspects are related through deterministic functions. Indeed, aspect \ref{aspnoe}
can be univocally determined from aspects \ref{asptr}, \ref{aspsoc} or \ref{aspcr}. Thus, the
information transmitted by aspect \ref{aspnoe} is also carried by any of the other aspects. In the
same manner, aspect \ref{aspsoc} can be determined from \ref{aspcr}. However, in \ref{appxB} we
prove that the number of patterns in the window (aspect \ref{aspnoe}) makes a vanishing
contribution to the information rate. That is, although aspect \ref{aspnoe} might be informative
for a finite window of length $\tau$, its contribution becomes negligible in the limit of long
windows. Surprisingly, the unordered set of pattern categories (aspect \ref{aspsoc}) also makes no
contribution to the information rate, as shown in \ref{appxC}. Even more, the entropy rates of both
aspects tend to zero in the limit of long time windows. Therefore, their information rate with
respect to any other aspect, of either the stimulus and/or the neural response, vanishes as the
window size increases. We thus do not discuss aspects \ref{aspnoe} and \ref{aspsoc} any further.

This is not the case of response aspects \ref{asptr} and \ref{aspcr}. In other words, they may
sometimes be informative; their definitions do not constrain them to be non-informative. Therefore,
in what follows, we transform the pattern representation into two other representations preserving
the precise timing of each pattern (the time representation) and the temporally-ordered pattern
categories (the category representation). Our goal is to determine in which way the precise timing
of each pattern conveys information about the time positions of stimulus features (the
\emph{when}), and how the temporally ordered pattern categories provide information about the
identity of the stimulus features (the \emph{what}).

\subsection{Time and category information}\label{results:tcinfo}

We define the \emph{time information} $I(\mbf{S};\mbf{T})$ as the mutual information rate between
the stimulus $\mbf{S}$ and the time representation $\mbf{T}$. In addition, we define the
\emph{category information} $I(\mbf{S};\mbf{C})$ as the mutual information rate between the
stimulus $\mbf{S}$ and the category representation $\mbf{C}$. The category information is novel
and, unlike and complementing previous studies \citep{gaudry2008, eyherabide2009}, allows us to
address the relevance of pattern categories in the neural code (see \ref{results:relevance}). Since
both $\mbf{T}$ and $\mbf{C}$ are transformations of the pattern representation $\mbf{B}$ (see Eqs.
\ref{representations}), the time and category information cannot be greater than the pattern
information, i.e.

\begin{equation}
 \left.
\begin{array}{l}
I(\mbf{S};\mbf{T})\\
I(\mbf{S};\mbf{C})
\end{array}
\right\} \leq I(\mbf{S};\mbf{B}) \, .
\end{equation}

\noindent When $\mbf{T}$ and $\mbf{C}$ are read out simultaneously, the pair $(\mbf{T}, \mbf{C})$
carries the same information as the pattern sequence $\mbf{B}$ ($I(\mbf{T}, \mbf{C};
\mbf{S})=I(\mbf{B}; \mbf{S})$). In fact, $\mbf{B}$ and the pair $(\mbf{T}, \mbf{C})$ are related
through a bijective function. To prove this, consider any pattern representation $\mbf{B}_i$ of a
neural response $\mbf{U}_i$. The pair $(\mbf{T}_i, \mbf{C}_i)$ associated with $\mbf{U}_i$ is a
function of $\mbf{B}_i$ (see Eqs.~\ref{representations}). Conversely, given the pair $(\mbf{T}_i,
\mbf{C}_i)$ associated with $\mbf{U}_i$, all the information about the time positions and
categories of patterns present in $\mbf{U}_i$ is available, and thus $\mbf{B}_i$ is univocally
determined. Notice that the pairs $(\mbf{T}, \mbf{C})$ are a subset of the Cartesian product $\mbf{T} \times \mbf{C}$.

The time positions of patterns may depend on their categories, and vice versa. To explore this
relationship, and how it affects the transmitted information, we separate the pattern information
as

\begin{equation}
I(\mbf{B};\mbf{S})=I(\mbf{S};\mbf{T})+I(\mbf{S};\mbf{C})+\Delta_{SR} \, ; \label{infoseparation}
\end{equation}

\noindent where $\Delta_{SR}$ represents the synergy/redundancy between the time and the category
representations, defined by

\begin{equation}
\Delta_{SR} = -I(\mbf{S};\mbf{T};\mbf{C})\, . \label{SynRed}
\end{equation}

\noindent Here, $I(X;Y;Z)=I(X;Y)-I(X;Y|Z)$ is called \emph{triple mutual information}
\citep{coverthomas, tsujishita1995}. If $\Delta_{SR}$ is positive, time and category information
are synergistic: more information is available when $\mbf{T}$ and $\mbf{C}$ are read out
simultaneously. Conversely, if $\Delta_{SR}$ is negative, time and category information are
redundant. The proof of Eq.~\ref{infoseparation} and Eq.~\ref{SynRed} is shown in \ref{appxD}.
Previous studies have already defined the synergy/redundancy for populations of neurons
\citep{schneidman2003}. It has also been applied to single neurons, to determine how different
aspects of response patterns encode the identity of single stimulus features \citep{furukawa2002,
nelken2005}. Here we extend the concept to encompass also dynamic stimuli where stimulus features
arrive at random times, as well as for arbitrary patterns, defined in time and/or across neurons.

As an example, consider the data presented in Figure~\ref{f4}, when the neural responses represented as a sequence of bursts ($\mbf{B}^{\alpha}$). For the case of the simulations (Figure~\ref{f4}.A), the time information is $I_{Sim}(\mbf{S}, \mbf{T}^{\alpha})= 180.4 \pm 0.2 \, bits/s$, and the category information,  $I_{Sim}(\mbf{S}, \mbf{C}^{\alpha})= 74.2 \pm 0.5 \, bits/s$. The synergy/redundancy term is slightly negative, but not significant ($\Delta_{SR}^{Sim}=-0.4 \pm 0.5$, two-sided t-test, $p(15)=0.44$). By construction, in the simulation the time and category information are neither redundant nor synergistic. For the experimental data (Figure~\ref{f4}.B), $I_{Exp}(\mbf{S}, \mbf{T}^{\alpha})= 63 \pm 2 \, bits/s$ and $I_{Exp}(\mbf{S}, \mbf{C}^{\alpha})= 50.6 \pm 0.6 \, bits/s$. In this case, we don't know whether the time information and the category information are redundant or synergistic before-hand. Yet, by comparing them with the pattern information we obtain $\Delta_{SR}^{Exp}= 7 \pm 3 bits/s$, indicating that timings and categories of patterns are slightly synergistic (two-sided t-test, $p(15)=0.063$).

The pattern, time and category information depend on the choice of the alphabet of patterns. For
example, the category information may increase or decrease depending on the nature of the aspect
defining the pattern categories \citep{furukawa2002, gollisch2008}. No general rules can be given,
predicting these changes: they depend on the neural representation at hand. However, when the
alternative pattern representations are linked through functions, some relations between their
variations can be predicted, without numerical calculations. Compare, for instance,
$\mbf{B}^{\alpha}$ and $\mbf{B}^{\beta}$ as defined in \ref{SpikePatternInfo}. By grouping all
bursts with more than one spike into a single category, not only $\mbf{B}^{\beta}$ is a function
$h_{\alpha \rightarrow \beta}$ of $\mbf{B}^{\alpha}$ ($\mbf{B}^{\beta}=h_{\alpha \rightarrow
\beta}(\mbf{B}^{\alpha})$), but also $\mbf{C}^{\beta}=h_{\alpha \rightarrow
\beta}(\mbf{C}^{\alpha})$. The time representation remains intact
($\mbf{T}^{\beta}=\mbf{T}^{\alpha}$). As a result, neither the pattern information nor the category
information can increase, whereas the time information remains constant. In addition, if
$\mbf{T}^{\alpha}$ and $\mbf{C}^{\alpha}$ are independent and conditionally independent given the
stimulus, so are $\mbf{T}^{\beta}$ and $\mbf{C}^{\beta}$. Therefore, the difference in the category
information equals the difference in the pattern information
($I(\mbf{S};\mbf{C}^{\alpha})-I(\mbf{S};\mbf{C}^{\beta})=I(\mbf{S};\mbf{B}^{\alpha})-I(\mbf{S};\mbf{B}^{\beta})$).

Analogously, consider a representation $\mbf{B}^{\gamma}$ in which the time positions of patterns
identified in $\mbf{B}^{\alpha}$ are read out with lower precision ($2\, \Delta t$). Since
$\mbf{B}^{\gamma}$ is a function of $\mbf{B}^{\alpha}$, two different responses
$\mbf{B}^{\alpha}_i$ and $\mbf{B}^{\alpha}_j$ that only differ little in the pattern time positions
are indistinguishable in the representation $\mbf{B}^{\gamma}$ ($\mbf{B}^{\gamma}_i =
\mbf{B}^{\gamma}_j$). In this case, the comparison between $\mbf{B}^{\alpha}$ and
$\mbf{B}^{\gamma}$ is analogous to the case analysed in the previous paragraph, with the role of
the time and category representations interchanged.

\begin{figure}[ht!]
\centering
\includegraphics{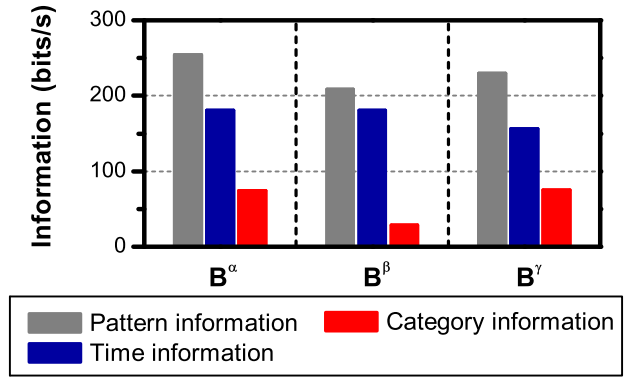}
\caption{\label{f6} {\bf Pattern, time and category information carried by different neural
representations.} The spike representation is transformed into a sequence of patterns:
\emph{$\mbf{B}^{\alpha}$ (left)}: grouped in categories according to the intra-pattern spike count;
\emph{$\mbf{B}^{\beta}$ (middle):} classified as isolated spikes or complex patterns; and
\emph{$\mbf{B}^{\gamma}$ (right):} classified as in $\mbf{B}^{\alpha}$, reading out the time
positions with a lower precision ($2\, \Delta t$). In all cases, error bars $< 1\%$. The simulation
data is taken from Figure~\ref{f4} (see \ref{detailsim} Simulation 1 for details).}
\end{figure}

We illustrate these results with an example. In Figure~\ref{f6}, the pattern, time and category
information are shown for three different choices of the pattern representation. The simulated
neural response is taken from Figure~\ref{f4}A. In the three cases, there is no synergy or
redundancy between the time and the category information ($\Delta_{SR}=0$). From Figure~\ref{f4}A,
we already know that $I(\mbf{S};\mbf{B}^{\beta})<I(\mbf{S};\mbf{B}^{\alpha})$. Comparing the left
and middle panels of Figure~\ref{f6}, we find that this reduction is due to a decrement in the
category information ($I(\mbf{S}, \mbf{C}^{\alpha})= 74.2 \pm 0.5 \, bits/s$, $I(\mbf{S},
\mbf{C}^{\beta})= 28.6 \pm 0.3 \, bits/s$, one-sided t-test, $p(10)<0.001$), as expected (see
\ref{SpikePatternInfo}). In agreement with the theoretical prediction, the time information remains
unchanged ($I(\mbf{S}, \mbf{T}^{\alpha})=I(\mbf{S}, \mbf{T}^{\beta})= 180.4 \pm 0.2 \, bits/s$,
one-sided t-test, $p(10)=0.5$).

Analogously, compare the left and right panels of Figure~\ref{f6}. In this case, both the pattern
and time information decrease ($I(\mbf{S}, \mbf{B}^{\alpha})= 254.2\pm 0.2\, bits/s$, $I(\mbf{S},
\mbf{B}^{\gamma})= 230.1 \pm 0.1\, bits/s$, $I(\mbf{S}, \mbf{T}^{\alpha})= 180.4 \pm 0.2\, bits/s$,
$I(\mbf{S}, \mbf{T}^{\gamma})= 156.0 \pm 0.2\, bits/s$, in both cases, one-sided t-test,
$p(10)<0.001$), while the category information remains unchanged ($I(\mbf{S}, \mbf{C}^{\alpha})=
I(\mbf{S}, \mbf{C}^{\gamma})= 74.2 \pm 0.5\, bits/s$, one-sided t-test, $p(10)=0.5$). Thus, as
mentioned previously, a reduction in the precision with which the patterns are read out always
decreases the time information, while keeping the category information constant.

In other examples, the variations in the time and category information may not be directly
accompanied by variations in the pattern information, due to the presence of synergy and
redundancy. For example, \citet{alitto2005} studied the encoding properties of tonic spikes,
long-ISI tonic spikes (tonic spikes preceded by long ISIs) and bursts. To evaluate the relevance of
distinguishing between tonic spikes and long-ISI tonic spikes, one can compare the information
conveyed by two representations: $\mbf{B}^{\xi}$, preserving the difference between tonic spikes
and long-ISI tonic spikes, and $\mbf{B}^{\phi}$, grouping them into the same category
\citep{gaudry2008}. Both $\mbf{B}^{\xi}$ and $\mbf{B}^{\phi}$ only differ in the category
representation, like $\mbf{B}^{\alpha}$ and $\mbf{B}^{\beta}$. However, unlike those
representations, $\Delta_{SR}^{\xi}$ and $\Delta_{SR}^{\phi}$ need not be either equal or zero, and
thus
$\left[I(\mbf{S};\mbf{B}^{\xi})-I(\mbf{S};\mbf{B}^{\phi})\right]=\left[I(\mbf{S};\mbf{C}^{\xi})-I(\mbf{S};\mbf{C}^{\phi})\right]+\left[\Delta_{SR}^{\xi}-\Delta_{SR}^{\phi}\right]$.
Indeed, by reading simultaneously the timing and category of a pattern, the uncertainty on whether
the following pattern will be a long-ISI tonic spike is reduced. Hence, this reduction is a source
of redundancy in $\mbf{B}^{\xi}$, where the long-ISI tonic spikes are explicitly identified. On the
other hand, the interpattern time interval (IPI) preceding a long-ISI tonic spike may reveal the
duration of the previous pattern. Any information contained in it constitutes a source of synergy
in $\mbf{B}^{\xi}$. The distinction between tonic spikes and bursts produces analogous effects on
the synergy and redundancy, affecting both representations $\mbf{B}^{\xi}$ and $\mbf{B}^{\phi}$.

As shown in \citet{coverthomas}, $I(\mbf{S};\mbf{T};\mbf{C})$ is symmetric in $\mbf{S}$, $\mbf{T}$
and $\mbf{C}$. Hence, $\Delta_{SR}$ is upper and lower bounded by

\begin{equation}\label{synredbounds}
-I(X;Y) \leq \Delta_{SR} \leq I(X;Y|Z)\, ;
\end{equation}

\noindent where $X$, $Y$ and $Z$ represent the variables $\mbf{S}$, $\mbf{T}$ and $\mbf{C}$ in such
an ordering that $I(X;Y)=\min\{I(\mbf{T};\mbf{C}), I(\mbf{S};\mbf{T}), I(\mbf{S};\mbf{C})\}$ (see
proof in \ref{appxE}). The same ordering applies for both bounds, in such a way that, for example,
if $I(\mbf{S};\mbf{T}|\mbf{C})$ is the least upper-bound, then $I(\mbf{S};\mbf{T})$ is the greatest
lower bound, from the set of bounds derived in Eq.~\ref{synredbounds}. These bounds are novel,
tighter than the bounds previously mentioned \citet{schneidman2003}.

If the left side of Eq. \ref{synredbounds} is zero, time and category information are non-redundant
($\Delta_{SR}\geq 0$). However, they may still be synergistic ($0 \leq \Delta_{SR}$), even in the
case when they are both zero ($I(\mbf{S};\mbf{T})=I(\mbf{S};\mbf{C})=0 \Rightarrow \Delta_{SR}\geq
0$). This property has often been overlooked \citep[see, for example,][]{foffani2009}. Time and
category information are non-synergistic if and only if the right side of Eq. \ref{synredbounds} is
zero. From the definition of the synergy/redundancy $\Delta_{SR}$ (Eq.~\ref{SynRed}), we show that

\begin{equation}\label{zerosynred}
\Delta_{SR}=0 \Leftrightarrow I(X;Y)=I(X;Y|Z)\, ;
\end{equation}

\noindent where $X$, $Y$ and $Z$ represent the variables $\mbf{S}$, $\mbf{T}$ and $\mbf{C}$ in any
order. In this case, the time and category information add up to the pattern information. This
situation may occur when either $I(X;Y)=I(X;Y|Z)\, =\, 0$ or $I(X;Y)=I(X;Y|Z)\, >\, 0$
\citep{schneidman2003,nirenberg2003}.

\subsection{Relevance and sufficiency of different aspects of the neural response}\label{results:relevance}

Previous studies have addressed the relevance of pattern timing in information transmission by
quantifying the time information and comparing it with the pattern information \citep{denning2005,
gaudry2008, eyherabide2009}. In other words, the relevance of pattern timing is given by the amount
of information carried by a representation that only preserves the time positions of patterns. We
call this paradigm \emph{criterium I}. Indeed, one can also address the relevance of pattern
categories using \emph{criterium I}. However, instead of quantifying the amount of information
carried by the category representation, these previous works have determined the information loss
due to ignoring the pattern categories. Here, this point of view is called \emph{criterium II}. In
what follows, we prove that \emph{criterium I} and \emph{criterium II} take into account different
information, and can thus lead to opposite results when both of them are applied to the same aspect
of the response.

Formally, under \emph{criterium I}, the pattern timing is relevant (or sufficient) for information
transmission if

\begin{equation}\label{critIT}
I(\mbf{S};\mbf{B})-\Delta I_{th}^I \leq I(\mbf{S};\mbf{T})\, .
\end{equation}

\noindent Here, $\Delta I_{th}^I$ represents a previously set threshold. Although
\citet{coverthomas} have defined sufficiency only for the case when $\Delta I_{th}^I=0$, in
practice, some amount of information loss ($\Delta I_{th}^I>0$) is usually accepted
\citep{nelken2007}. We can also employ this criterium to address the relevance of pattern
categories, comparing

\begin{equation}\label{critIC}
I(\mbf{S};\mbf{B})-\Delta I_{th}^I \leq I(\mbf{S};\mbf{C})\, .
\end{equation}

\noindent On the other hand, under \emph{criterium II}, the pattern categories are relevant to
information transmission if

\begin{equation}\label{critIIC}
I(\mbf{S};\mbf{T}) \leq I(\mbf{S};\mbf{B})-\Delta I_{th}^{II} \, .
\end{equation}

\noindent Therefore, pattern categories are relevant if pattern timings transmit little
information, irrespective of the information carried by categories themselves. Remarkably, if
$\Delta I_{th}^{I}=\Delta I_{th}^{II}$, the pattern categories are relevant (irrelevant) if and
only if the pattern timings are irrelevant (relevant) (compare Eqs.~\ref{critIT} and
\ref{critIIC}).

From the bijectivity between $\mbf{B}$ and $(\mbf{T};\mbf{C})$ (see \ref{results:tcinfo}), we find
that \emph{criterium II} can be written as

\begin{equation}\label{critIICnueva}
\Delta I_{th}^{II}-\Delta_{SR} \leq I(\mbf{S};\mbf{C}) \, .
\end{equation}

\noindent As a result, under \emph{criterium II}, the relevance of an aspect depends not only on
the information conveyed by that very aspect --- as in \emph{criterium I} --- but also on the
synergy/redundancy between that aspect and the complementary ones. Both criteria coincide when
$\Delta I_{th}^{I}+\Delta I_{th}^{II}=I(\mbf{S};\mbf{B})+\Delta_{SR}$ (compare Eqs.~\ref{critIC}
and \ref{critIICnueva}), implying that equality in the thresholds is neither necessary nor
sufficient to obtain a coincidence.

By using \emph{criterium I} for the relevance of pattern timing and \emph{criterium II} for the
relevance of pattern categories, the information that is repeated in both aspects (redundant
information) only contributes to the relevance of the pattern timing. However, the information that
is carried in both aspects simultaneously (synergistic information) only contributes to the
relevance of the pattern categories. The discrepancies in this way induced are shown in the
following example. Consider that $I(\mbf{S};\mbf{R})=10\, bits/s$, $I(\mbf{S};\mbf{T})=9\, bits/s$,
$I(\mbf{S};\mbf{C})=10\, bits/s$ and $\Delta I_{th}=2\, bits/s$. Under \emph{criterium II},
$\mbf{C}$ is irrelevant because $I(\mbf{S};\mbf{B})-I(\mbf{S};\mbf{T})=1\, bit/s <\Delta I_{th}$.
Nevertheless, under \emph{criterium I}, $\mbf{C}$ is necessarily relevant, since it constitutes a
sufficient statistics ($I(\mbf{S};\mbf{B})=I(\mbf{S};\mbf{C})$). Analogous results are obtained for
the relevance of pattern timing. In addition, different thresholds are used for the relevance of
each aspect (compare Eqs.~\ref{critIT} and \ref{critIIC}). In the previous example, the pattern
timing is relevant only if $I(\mbf{S};\mbf{T})>8 \, bits/s$ whereas the pattern categories are
relevant only if $I(\mbf{S};\mbf{C})> 2\, bits/s$, showing an unjustified asymmetry between both
aspects.

\subsection{Time and category entropy of the stimulus}\label{results:stimulusentropy}

Many studies have interpreted that pattern-based codes function as \emph{feature extractors}, where
the identity of each stimulus feature (the \emph{what}) is represented in the pattern category
$\mbf{C}$, and the timing of each stimulus feature (the \emph{when}), in the pattern temporal
reference $\mbf{T}$ (see \emph{Introduction} and references therein). To assess this standard view,
we formally define the \emph{what} and the \emph{when} in the stimulus, and relate them with the
time and category information. In the next subsection, we determine the conditions that are
necessary and sufficient for the standard view to hold. Finally, we show that small
category-dependent changes in the timing of patterns (such as latencies) may induce departures from
the standard view (altering both the amount and the composition of the information carried by
$\mbf{T}$ and $\mbf{C}$).

Since the stimulus $\mbf{S}$ is composed of discrete features (see \nameref{methods} for a
discussion on continuous stimuli), it can also be written in terms of a time ($\mbf{S_T}$) and a
category ($\mbf{S_C}$) representation, such that $\mbf{S}$ and the pair $(\mbf{S_T}, \mbf{S_C})$
are related through a bijective map. We formally define the \emph{what} in the stimulus as the
category representation $\mbf{S_C}$, and the \emph{when} as the time representation $\mbf{S_T}$.
Indeed, $\mbf{S_T}$ indicates \emph{when} the stimulus features occurred, whereas $\mbf{S_C}$ tags
\emph{what} features appeared.

The \emph{stimulus entropy} is defined as the entropy rate $H(\mbf{S})$, while the stimulus
\emph{time entropy} and \emph{category entropy} are the entropy rates $H(\mbf{S_T})$ and
$H(\mbf{S_C})$, respectively. The time and category entropies are intimately related to \emph{when}
and \emph{what} features happened: they are a measure of the variability in the time positions and
categories of stimulus features, respectively. These quantities were previously defined for Poisson
stimuli in \citet{eyherabide2010}, and here these definitions are generalised to encompass any
stochastic stimulus. Since $\mbf{S}$ and $(\mbf{S_T}, \mbf{S_C})$ are related through a bijective
function,

\begin{equation}
H(\mbf{S})=H(\mbf{S_T})+H(\mbf{S_C})-I(\mbf{S_T}, \mbf{S_C})\, ;
\end{equation}

\noindent where the information rate $I(\mbf{S_T}, \mbf{S_C})$ is a measure of the redundancy
between the time and category entropies of the stimulus. Since $I(\mbf{S_T}, \mbf{S_C})$ is always
non-negative, $\mbf{S_T}$ and $\mbf{S_C}$ cannot be synergistic.

The standard view of the role of patterns formally implies that the category information
$I(\mbf{S}, \mbf{C})$ (the time information $I(\mbf{S}, \mbf{T})$) can be reduced to the mutual
information $I(\mbf{S_C}, \mbf{C})$ ($I(\mbf{S_T}, \mbf{T})$). Therefore, $H(\mbf{S_C})$ and
$H(\mbf{S_T})$ must be upper bounds for the category and time information, respectively. However,
these bounds are not guaranteed by the mere presence of patterns in the neural response. Some cases
may be more complicated because, for example, $\mbf{S_C}$ and $\mbf{S_T}$ may not be independent
variables (see \ref{methods:simdata}). A dependency between these two stimulus properties implies
that the \emph{what} and the \emph{when} are not separable concepts.

\subsection{The canonical feature extractor}\label{result:ife}

In this section, we determine the conditions under which the standard interpretation holds: The
category information represents the knowledge on the \emph{what} in the stimulus, and the time
information, the knowledge on the \emph{when}. To that aim, we define a \emph{canonical feature extractor} as a neuron model in which

\begin{subequations} \label{condideal}
\begin{align}
I(\mbf{T};\mbf{S_C}|\mbf{S_T})&=0 \label{infotempideal}\\
I(\mbf{C};\mbf{S_T}|\mbf{S_C})&=0 \label{infocatideal}\, .
\end{align}
\end{subequations}

\noindent Under each of these conditions, the time and category information become

\begin{subequations}\label{infoidealbounds}
\begin{align}
I(\mbf{T};\mbf{S})&=I(\mbf{T};\mbf{S_T})\leq H(\mbf{S_T})\\
I(\mbf{C};\mbf{S})&=I(\mbf{C};\mbf{S_C})\leq H(\mbf{S_C})\, .
\end{align}
\end{subequations}

\noindent Consequently, the response pattern categories represent \emph{what} stimulus features are
encoded, whereas the pattern time positions represent \emph{when} the stimulus features occur. In
particular, the time and category information are upper bounded by the stimulus time and category
entropies, respectively.

Condition \ref{infotempideal} implies that all the information $I(\mbf{S_C};\mbf{T})$ is already
contained in the information $I(\mbf{S_T};\mbf{T})$. In other words, $I(\mbf{S_C};\mbf{T})$ is
completely redundant with $I(\mbf{S_T};\mbf{T})$, and $I(\mbf{S_C};\mbf{T}) \leq
I(\mbf{S_T};\mbf{T})$. In this sense, we say that the time information represents the \emph{when}
in the stimulus. Analogous implications can be obtained from condition \ref{infocatideal} for the
category representation $\mbf{C}$, by interchanging $\mbf{T}$ with $\mbf{C}$, and $\mbf{S_T}$ with
$\mbf{S_C}$ (see formal proof in \ref{appxF}). Therefore, conditions \ref{condideal} are necessary
and sufficient to ensure that the standard view of the role of patterns in the neural code actually
holds (see \ref{whatwhen}).

A canonical feature extractor does not require $\mbf{T}$ and $\mbf{C}$ to be independent nor
conditionally independent given the stimulus. In other words, the time and category information may
or may not be synergistic or redundant, and the timing (category) of each individual pattern may or
may not be correlated with other pattern time positions (pattern categories) or even with pattern
categories (pattern time positions). In addition, conditions~\ref{condideal} may also encompass
situations in which some information about $\mbf{S_C}$ ($\mbf{S_T}$) is carried by $\mbf{T}$
($\mbf{C}$), but not by $\mbf{C}$ ($\mbf{T}$).

In order to see how synergy and redundancy behave in a canonical feature extractor, we replace
Eqs.\ref{infoidealbounds} in Eq. \ref{infoseparation}, and obtain

\begin{equation}
I(\mbf{B};\mbf{S})=I(\mbf{T};\mbf{S_T})+I(\mbf{C};\mbf{S_C})+\Delta_{SR}\, .\label{idealtimecatsep}
\end{equation}

\noindent We find that, for a canonical feature extractor, the synergy/redundancy $\Delta_{SR}$ is
lower bounded by

\begin{equation}
-I(\mbf{S_T};\mbf{S_C})\leq \Delta_{SR}\, ; \label{boundsynred}
\end{equation}

\noindent (see proof in \ref{appxG}). In other words, the synergy/redundancy term $\Delta_{SR}$
cannot be smaller than the redundancy --- already present in the stimulus --- between the timing
and categories of stimulus features. In addition, the absence of redundancy in the stimulus
($I(\mbf{S_T};\mbf{S_C})=0$) constrains the neural model to be non-redundant ($\Delta_{SR}\geq 0$).

Consider a neural model in which $\mbf{T}=f(\mbf{S_T};\psi_T)$ and $\mbf{T}=f(\mbf{S_C};\psi_C)$,
where $\psi_T$ and $\psi_C$ are independent sources of noise, such that $p(\psi_T, \psi_C,
\mbf{S_T}, \mbf{S_C})=p(\psi_T)\, p(\psi_C) \, p(\mbf{S_T}, \mbf{S_C})$. Thus, $\mbf{T}$ and
$\mbf{C}$ are two channels of information under independent noise \citep{shannon1948, coverthomas}.
This model constitutes a canonical feature extractor. Indeed, $\mbf{T}$ ($\mbf{C}$) is only related to
$\mbf{S_C}$ ($\mbf{S_T}$) through $\mbf{S_T}$ ($\mbf{S_C}$), thus complying with
condition~\ref{infotempideal} (~\ref{infocatideal}). In addition, if $\mbf{S_T}$ and $\mbf{S_C}$
are independent, then $\mbf{T}$ and $\mbf{C}$ constitute independent channels of information
\citep{coverthomas, gawne1993}. This model plays a prominent role in the interpretation of neurons
and neural pathways as channels of information \citep{gawne1993, schneidman2003, montemurro2008,
krieghoff2009}, as discussed in \ref{encdec}.

\begin{figure}[ht]
\centering
\includegraphics{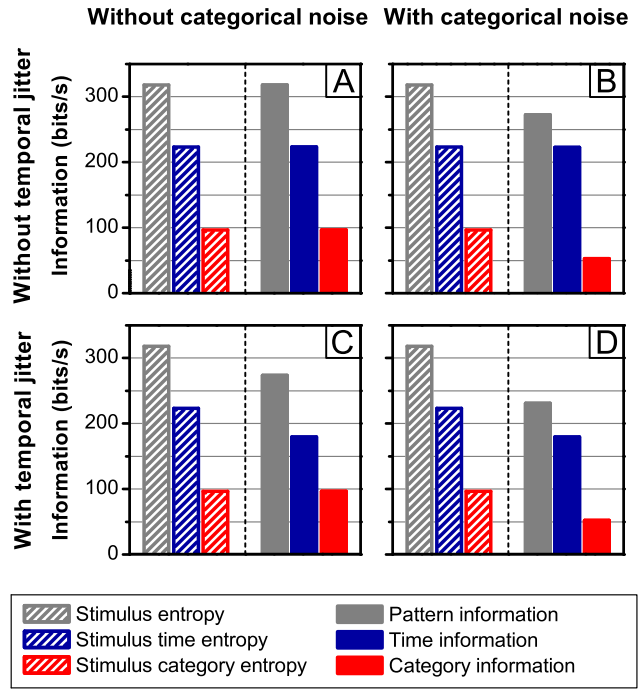}
\caption{\label{f7} {\bf Information transmitted by a canonical feature extractor under different
noise conditions.} The left side of each panel shows the stimulus entropy, whereas the right side
shows the pattern, time and category information. In all cases, $\Delta_{SR}=0$, so the pattern
information is equal to the sum of the category and the time information. \emph{From left to
right}: The addition of categorical noise reduces only the category information irrespective of the
amount of temporal jitter. \emph{From top to bottom}: The presence of temporal jitter degrades
solely the time information irrespective of the amount of categorical noise. The pattern
information is upper bounded by the stimulus entropy, the time information by the stimulus time
entropy, and the category information by the stimulus category entropy. In all cases, error bars $<
1\%$. For detailed description of the simulation see \ref{detailsim} Simulation 2.}
\end{figure}

The independent channels of information may be regarded as the simplest canonical feature extractor.
Since $\mbf{T}$ and $\mbf{C}$ are independent and conditionally independent given $\mbf{S}$, the
time and category information add up to the pattern information ($\Delta_{SR}=0$). An example of
this model is shown in Figure~\ref{f7}. In the four simulations carried out, the neural responses
consist of a sequence of four different patterns, associated with four different stimulus features,
under the presence or absence of temporal jitter and categorical noise (see \ref{detailsim}
Simulation 2 for a detailed description; Figure \ref{f2}B shows examples of the different noise
conditions). In Figure~\ref{f7}, the spike information is omitted because it coincides with the
pattern information (all cases, one-sided t-test, $p(10)=0.5$). Indeed, by construction, all the
information is transmitted by patterns, which can be univocally identified in the response. In
agreement with the theoretical results (Eq. \ref{infoidealbounds}), the time and the category
information are always upper-bounded by the stimulus time and category entropy, respectively (all
cases, one-sided t-test, $p(10)>0.4$).

Comparing upper and lower panels of Figure~\ref{f7}, we show that the time information is degraded
by the addition of temporal jitter (both cases, one-sided t-test, $p(10)<0.001$), while the
category information remains constant (both cases, one-sided t-test, $p(10)>0.14$). Analogously,
comparing left and right panels of Figure~\ref{f7}, we find that the addition of categorical noise
decreases the category information (both cases, one-sided t-test, $p(10)<0.001$), while keeping the
time information constant (panel A and B, $I(\mbf{S};\mbf{T}^A)=223.3 \pm 0.1\, bits/s$,
$I(\mbf{S};\mbf{T}^B)=222.8 \pm 0.1\, bits/s$, one-sided t-test, $p(10)=0.08$; panel C and D,
one-sided t-test, $p(10)=0.5$). This is expected since, by construction, the categorical noise only
depends on the stimulus categories and affects solely the pattern categories, whereas the temporal
jitter considered here only affects the pattern time positions, irrespective of their categories or
the stimulus.

\subsection{Departures from the canonical feature extractor}\label{departures}

The example shown in Figure~\ref{f7} turns out to be more complicated if the pattern timing depends
on the pattern category, as occurs in latency codes \citep{gawne1996, furukawa2002, chase2007,
gollisch2008}. Indeed, in those cases, the comparison between the timing of response patterns and
the timing of stimulus features carries information about the stimulus categories
($I(\mbf{S_C};\mbf{T}|\mbf{S_T})>0$). As a result, Eq.~\ref{infotempideal} does not hold. Latency
codes may be an intrinsic property of the encoding neuron, may result as a consequence of synaptic
transmission \citep{lisman1997, reinagel1999}, or may either arise from the convention used to
construct the pattern representation, for example, ascribing the timing of a pattern as the mean
response time, the first or any other spike inside the pattern \citep{nelken2005, eyherabide2008}.
In all these cases, a latency-like dependence between the time positions and categories of patterns
may arise.

To assess the effect of different latencies associated with each pattern category on the neural
response, consider the neural model used in Figure~\ref{f7}, except that now, the pattern latencies
vary with the pattern category $b$, according to $\mu_b=1+\alpha_{\mu}*(4-b)$. Here $\alpha_{\mu}$
is the \emph{latency index}, representing the difference between the latencies of consecutive
pattern categories. Three values of $\alpha_{\mu}$ were considered: $0$, $2$ and $4\, ms$. When
$\alpha_{\mu}=0\, ms$, all patterns have the same latencies. This case was analysed in Figure
\ref{f7}. As $\alpha_{\mu}$ increases, so does the latency difference of different patterns.

Due to the deterministic link between the pattern latencies and pattern categories, the pattern
representations ($\mbf{B}^0$, $\mbf{B}^2$ and $\mbf{B}^4$), associated with the different values of
$\alpha_{\mu}$ are related bijectively. In addition, the category representation does not depend on
$\alpha_{\mu}$. Only the time representation is altered by a change in the latency index,
irrespective of the presence of absence of temporal jitter and categorical noise. Therefore, any
change in the time information is immediately reflected in the synergy/redundancy term

\begin{subequations}\label{synreddeparture}
\begin{align}
\Delta_{SR}^{x}&=I(\mbf{S};\mbf{T}^{0})-I(\mbf{S};\mbf{T}^{x})\, . \\
&= -\left[H(\mbf{T}^{x})-H(\mbf{T}^{0})\right] +
\left[H(\mbf{T}^{x}|\mbf{S})-H(\mbf{T}^{0}|\mbf{S})\right]\, .
\end{align}
\end{subequations}

\noindent Here, $\Delta_{SR}^{x}$ and $\mbf{T}^{x}$ represent the synergy/redundancy term and the
time representation, respectively, for $\alpha_{\mu}=x\, ms$.

The impact of different latencies is twofold. In the first place, the presence of categorical noise
increments the temporal noise through the deterministic link between latencies and categories.
Therefore, the time noise entropy (time information) when $\alpha_{\mu}>0$ is greater (less) than
that when $\alpha_{\mu}=0$. However, this does not occur when the time and category representations
are read out simultaneously. Indeed, given the category representation, any time representation for
$\alpha_{\mu}=x>0$ can be univocally determined from the time representation for $\alpha_{\mu}=0$,
and vice versa, counteracting the effect of the temporal noise. Therefore, the variation in the
time noise entropy ($H(\mbf{T}^{x}|\mbf{S})-H(\mbf{T}^{0}|\mbf{S})$ in Eq.~\ref{synreddeparture})
can be regarded as a source of synergy.

In the second place, the variation in the latencies modifies the inter-pattern time interval
distribution, incrementing the time total entropy (and the time information) when $\alpha_{\mu}>0$
with respect to the case when $\alpha_{\mu}=0$. In addition, this variation introduces information
about the pattern categories in the inter-pattern time interval, and consequently it also
introduces information about the stimulus identities. For example, a short interval between two
consecutive patterns indicates that the second patterns belongs to a category with a short latency.
In consequence, the increment in the time total entropy ($H(\mbf{T}^{x})-H(\mbf{T}^{0})$ in
Eq.~\ref{synreddeparture}) can be regarded as a source of redundancy.

\begin{figure}[htp]
\includegraphics{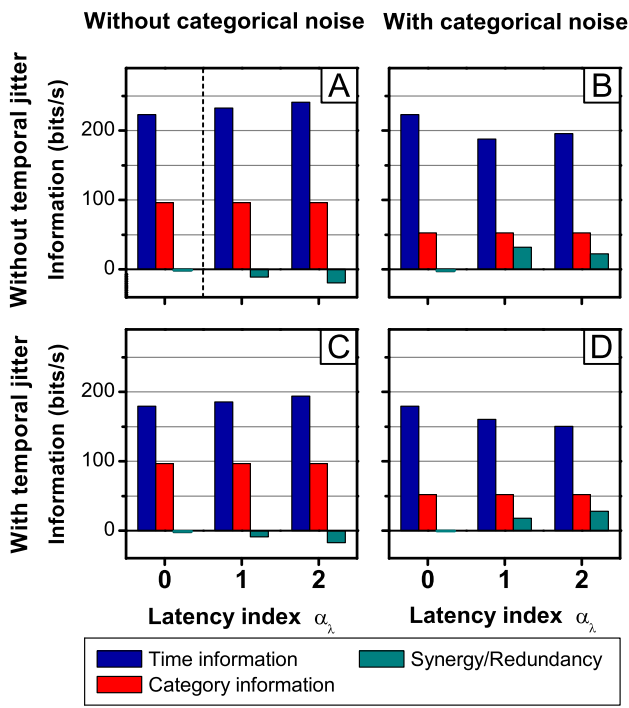}
\caption{\label{f8} {\bf Examples of departures from the behaviour of the canonical feature extractor:
The effect of pattern-category dependent latencies.} In all cases, when latencies depend on the
pattern category, the time information is affected while the category information remains
unchanged. Furthermore, the addition of categorical noise not only affects the category information
but also the time information. In general, how the addition of temporal and/or categorical noise
affects the time information depends on the latency index, as well as on the noise already present
in the response. For simulation details, see \ref{detailsim} Simulation 2. The case where
$\alpha_{\mu}=0\, ms$ was analysed in Figure~\ref{f7} and is reproduced here for comparison. In
panel {\bf A}, the case where $\alpha_{\mu}=0\, ms$ also represents the stimulus entropies, as
shown in Figure~\ref{f7}A. In all cases, error bars $< 1\%$.}
\end{figure}

To illustrate these theoretical inferences, the results of the simulations are shown in
Figure~\ref{f8}. As expected, when $\alpha_{\mu}=x>0$, the latencies alter the time information.
However, they do not alter the pattern nor the category information, and thus any variation in the
time information is compensated by an opposite variation in the synergy/redundancy term. Notice
that the changes in the time information not only depend on the latency index, but also on the
presence of temporal and categorical noise. Indeed, in the absence of categorical noise,
$H(\mbf{T}^{x}|\mbf{S})=H(\mbf{T}^{0}|\mbf{S})=0$, and thus $\Delta_{SR}\leq 0$. The effect of the
temporal jitter depends on its distribution as well as the distribution of the inter-pattern time
intervals, so this analysis if left for future work.

In these examples we see that for non-canonical feature extractors, one can no longer say that the
pattern categories represent the \emph{what} in the stimulus and the pattern timings represent the
\emph{when}, not even in the absence of synergy/redundancy. As shown in Eq.~\ref{synreddeparture},
$\Delta_{SR}$ results from a complex tradeoff between the effect of categorical noise on the total
and noise time response entropies. This tradeoff depends on the latency index and the amount of
temporal noise in the system, as shown in Figure~\ref{f8}.

Latency-like effects may be involved in a translation from a pattern-duration code into an
inter-spike interval code \citep{reich2000, denning2005}. Indeed, bursts may increase the
reliability of synaptic transmission \citep{lisman1997}, making it more probable to occur at the
end of the burst. In that case, the duration of the burst determines the latency of the
postsynaptic firing. In particular, this indicates that bursts can be simultaneously involved in
noise filtering and stimulus encoding, in spite of the belief that these two functions cannot
coexist \citep{krahe2004}. Notice that here, latency codes have been studied for well-separated
stimuli. However, if patterns are elicited close enough in time, they may interfere in a diversity
of manners \citep{fellous2004}, precluding the code from being read out. Although we cannot address
all these cases in all generality, the framework proposed here is valid to address each particular
case.

\section{Discussion}

In this paper, we have focused on the analysis of temporal and categorical aspects, both in the
stimulus and the response. Our results, however, are also applicable to other aspects. In the case
of responses, these aspects can be latencies, spike counts, spike timing variability,
autocorrelations, etc. Examples of stimulus aspects are colour, contrast, orientation, shape,
pitch, position, etc. The only requirement is that the considered aspects be obtained as
transformations of the original representation, as defined in \ref{methods:representations} (see
\ref{difapp}). The information transmitted by generic aspects can be analysed by replacing
$\mbf{B}$ ($\mbf{S}$) with a vector representing the selected response (stimulus) aspects. The
amount of synergy/redundancy between aspects is obtained from the comparison between the
simultaneous and individual readings of the aspects. In addition, the results can be generalised
for aspects defined as statistical (that is, non-deterministic) transformations of the neural
response, or of the stimulus. The data processing inequality also holds in those cases
\citep{coverthomas}.

\subsection{Meaning of time and category information and their relation with the \emph{what} and the \emph{when} in the
stimulus} \label{whatwhen}

\begin{figure}[htp]
\centering
\includegraphics{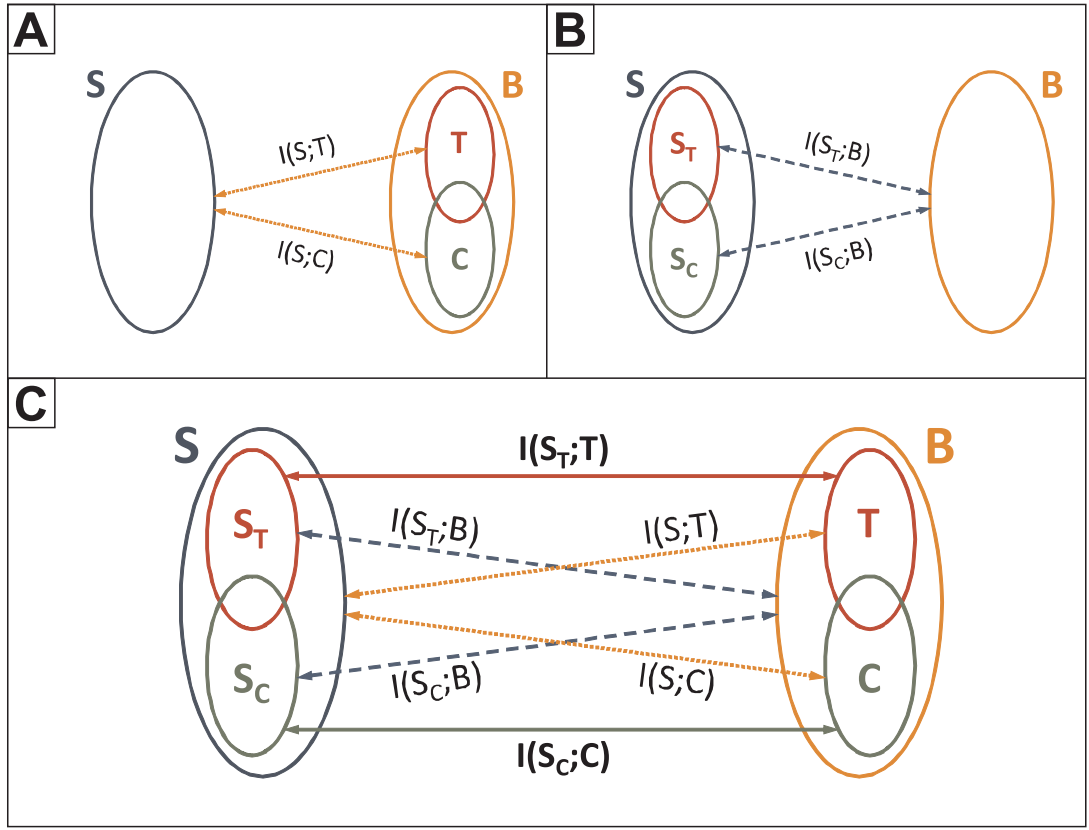} \caption{\label{f9}
{\bf Analysis of the role of spike patterns: Relationship with the \emph{what} and the \emph{when}
in the stimulus.} ({\bf A}) Categorical and temporal aspects in the neural response. Definitions of
time $I(\mbf{S};\mbf{T})$ and category $I(\mbf{S};\mbf{C})$ information. ({\bf B}) Categorical and
temporal aspects in the stimulus. Information about the \emph{what} $I(\mbf{S_C};\mbf{B})$ and the
\emph{when} $I(\mbf{S_T};\mbf{B})$ conveyed by the neural response $\mbf{B}$. ({\bf C}) Analysis of
the role of patterns in the neural response. Mutual information between different aspects of the
stimulus and different aspects of the neural response.}
\end{figure}

In this paper, we defined the category and the time information in terms of properties of the
neural response. The category (time) information is the mutual information between the \emph{whole
stimulus} $\mbf{S}$ and the categories $\mbf{C}$ (timing $\mbf{T}$) of response patterns (see
Figure~\ref{f9}A). These definitions only require the neural response to be structured in patterns.
No requirement is imposed on the stimulus, i.e. the stimulus need not be divided into features. Our
definitions, hence, are not symmetric in the stimulus and the response. In some cases, however, the
stimulus is indeed structured as a sequence of features. One may ask how the stimulus identity (the
\emph{what}) and timing (the \emph{when}) is encoded in the neural response (see Figure~\ref{f9}B).
To that end, we defined the \emph{what} in the stimulus in terms of the category representation
($\mbf{S_C}$), and the \emph{when}, in terms of the time representation ($\mbf{S_T}$).

These rigorous definitions allowed us to disentangle how the \emph{what} and the \emph{when} in the
stimulus are encoded in the category and time representations of the neural response. We calculated the mutual information rates between different aspects of the stimulus and different aspects of the neural response (see Figure~\ref{f9}C). In the standard view, the pattern categories are assumed to encode the \emph{what} in the stimulus, and the timing of patterns, the \emph{when} \citep{theunissen1995, borst1999, martinezconde2002, krahe2004, alitto2005, oswald2007, eyherabide2008}. These assumptions have been stated in qualitative terms. There are two different ways in which the standard view can be formalized as a precise assertion.

On one hand, the standard view can be seen as the assumption that
the category (time) representation only conveys information about
the \emph{what} (the \emph{when}). Evaluating this assumption involves the
comparison between the
information conveyed by the category (time) representation about the whole stimulus (dotted lines
in Figure~\ref{f9}C) with the information that this same representation conveys about the
\emph{what} (the \emph{when}) in the stimulus (solid lines in Figure~\ref{f9}C). Formally, this
means to address whether $I(\mbf{S};\mbf{C})=I(\mbf{S_C};\mbf{C})$ (whether
$I(\mbf{S};\mbf{T})=I(\mbf{S_T};\mbf{T})$). In this sense, we say that the category (time)
information only represents the \emph{what} (the \emph{when}) in the stimulus. A system complying with this first interpretation of the standard view was
called a canonical feature extractor (see \ref{result:ife}).

On the other hand, the second way to define the standard view
rigorously is to assume that the \emph{what} (the \emph{when}) is completely
encoded by the category (time) representation. Testing this second
assumption involves the comparison between the
information about the \emph{what} (the \emph{when}), conveyed by the category (time) representation
(solid lines in Figure~\ref{f9}C) and by the pattern representation of the neural response (dashed
lines in Figure~\ref{f9}C). Formally, it involves assessing whether
$I(\mbf{S_C};\mbf{B})=I(\mbf{S_C};\mbf{C})$ (whether $I(\mbf{S_T};\mbf{B})=I(\mbf{S_T};\mbf{T})$).
In this sense, we say that all the information about the \emph{what} (the \emph{when}) in the
stimulus is encoded in the category (time) representation of the neural response. A system for
which these equalities hold is called a canonical feature interpreter. It is analogous to the canonical feature extractor, with the role of the stimulus and the response interchanged (see \ref{appxH}).

The two formalizations of the standard view are complementary. The first one assesses how different aspects of the stimulus
are encoded in each aspect on the neural response. The second one focuses on how each aspect of the
stimulus is encoded in different aspects of the neural response. Thus, the second approach is a
symmetric version of the first one. However, a canonical feature extractor might or might not be a
canonical feature interpreter, and vice versa. A perfect correspondence between the \emph{what} and the
\emph{when} on one side, and pattern timing and categories, on the other, is found for systems that
are canonical feature extractors and canonical feature interpreters, simultaneously.

\subsection{Two different approaches to the analysis of neural codes}\label{difapp}

In order to understand a neural code, one needs to identify those aspects of the neural response
that are relevant to information transmission. To that aim, two different paradigms have been used:
\emph{criterium I}, assessing the information that one aspect conveys about the stimulus, and
\emph{criterium II}, assessing the information loss due to ignoring that aspect (see
\ref{results:relevance}). Previous studies have used \emph{criterium I} to analyse the relevance of
spike counts \citep{furukawa2002, foffani2009}, spike patterns \citep{reinagel1999,
eyherabide2008}, and pattern timing \citep{denning2005, gaudry2008, eyherabide2009}. However, when
assessing the relevance of the complementary aspects, such as spike timing and internal structure
of patterns, these studies have used \emph{criterium II}. As a result, in these studies the
relevance of the tested aspect is conditioned to the irrelevance of the other aspects.

There are cases where building a representation that preserves a definite response aspect is not
evident (nor perhaps possible). Such is the case, for example, when assessing the differential
roles of spike timing and spike count: It is not possible to build a representation preserving the timing of the spikes without preserving the spike count (see \ref{results:infoaspects}). It is instead possible to only preserve the spike count. Since the
spike-count representation is a function of the spike-timing representation, one may argue that
there is an intrinsic hierarchy between the two aspects. The same situation is encountered  when
evaluating the information encoded by the pattern representation, as compared to the spike
representation (see \ref{SpikePatternInfo}). There, it was not possible to construct a
representation only containing those aspects that had been discarded in the pattern representation.
However, this is not the case when evaluating the differential role between pattern timing and
pattern categories, or the relevance of a specific pattern category.

In the present study, we take advantage of both approaches. Firstly, we notice that pattern timing
and pattern categories are complementary response aspects, and quantify the information preserved
by each aspect (see \ref{results:tcinfo}). Then, we determine whether there is synergy or
redundancy between the time and category information, which is formally equivalent to comparing the
information preserved by (\emph{criterium I}) and lost due to ignoring (\emph{criterium II}) each of
the two aspects. As a result, we gain insight on the relevance of each aspect as well as how the
aspects interact to transmit information (see \ref{discsynergy}). These procedures can be extended
to encompass any two different aspects of the neural response (see \ref{encdec}).

Notice that the role of correlations, both in time and/or across neurons, has been evaluated using
\emph{criterium II} \citep{brenner2000, dayan2001, nirenberg2001, petersen2002, schneidman2003,
montemurro2007b}. However, these authors did not build two complementary representations of the
neural response ignoring and preserving the correlations, as proposed here. Instead, they ignored
correlations by constructing artificial neural responses (or artificial response probabilities)
where different neurons were independent or conditionally independent. Thus, their analysis
involves a comparison between the real and the artificial neural code. Our analysis, instead, is
completely based on complementary reductions of the real neural response. Moreover, in previous
studies, the artificial neural responses are not a transformed version of the real response in a well defined time window. Thus, in some cases, the difference between the information with and without preserving correlations is not guaranteed to be
non-negative by the \emph{data processing inequality} \citep{coverthomas}.

\subsection{Representations of the neural response and the data processing
inequality}\label{dpidiscuss}

In some previous studies, the information encoded by different response aspects was assessed, as here, by transforming each neural response window ($\mbf{R}_{\tau}$) of size $\tau$
through functions, into the pattern representation ($\mbf{B}_{\tau}$) \citep{furukawa2002,
petersen2002, nelken2005, gollisch2008}. Examples of those response aspects are the first-spike
latencies, spike counts, spike-timing variabilities and first (second, third, etc.) spikes in a
pattern. As a result, the information carried by the individual response aspects cannot be greater
than that provided by the neural response in the same window ($I(\mbf{S};\mbf{B}_{\tau})\leq
I(\mbf{S};\mbf{R}_{\tau})$), irrespective of the length $\tau$ (see \ref{infocalculo}). In other studies, however, the pattern representation $\mathcal{B}$ was obtained by transforming the spike representation inside a sliding window of
variable length: the length of the window depended on the category of the actual pattern. Then, $\mathcal{B}$ was read out with time windows of size $\tau$. That is the
case, for example, when addressing the information conveyed by inter-spike intervals of length
$>38\, ms$ using words of length $\tau=14.8\, ms$ \citep{reich2000} and by patterns of length
$>104\, ms$, $>10\, ms$ and $>56\, ms$, using time windows up to $64\, ms$, $3.2\, ms$ and $16\,
ms$, respectively \citep{reinagel2000, eyherabide2008, gaudry2008}. Unlike the first approach, in
this case the \emph{data processing inequality} does not apply, since $\mathcal{B}_{\tau}$ is not a function of
$\mbf{R}_{\tau}$. Therefore, $I(\mbf{S};\mbf{B}_{\tau})$ can be larger or smaller than $I(\mbf{S};\mbf{R}_{\tau})$. However, when $\tau \rightarrow \infty$, $\mathcal{B}_{\tau}= \mbf{B}_{\tau}$, so
asymptotically, both approaches coincide.

\subsection{The role of synergy and redundancy in the search for relevant response aspects} \label{discsynergy}

One of the main goals of the analysis of the neural code is to identify the response aspects that
are relevant to information transmission. In this context, two important questions arise: how
relevant the chosen aspects are, and how autonomously they stand. Their relevance to information
transmission is assessed with information-theoretical measures, as exemplified here with the
category and time information (see \ref{difapp}). Their autonomy refers to whether each aspect transmits information
by itself or not, and whether the transmitted information is
shared by other aspects or not. The degree of autonomy is assessed by quantifying the synergy/redundancy term
($\Delta_{SR}$) between the different aspects.

The concept of synergy/redundancy entails the comparison between the effect of the whole and the
sum of the individual effects of the constituent parts. The concept requires the constituent parts to be univocally determined by the whole, as well as the whole to be completely determined given its constituent parts. In other
words, the whole and the constituent parts must be related through a bijective function. In neuroscience, the synergy/redundancy between groups of neurons has been addressed by comparing the information carried by the group of
neurons (the whole) and the sum of the information of each and every neuron from the group (the
constituent parts) \citep{brenner2000,schneidman2003}. As a result, $\Delta_{SR}$ can be interpreted as a trade-off between synergy and redundancy \citep{schneidman2003}.

Intuitively, the presence of synergy ($\Delta_{SR}>0$) between two aspects indicates that, for many
responses, the aspects must be read out simultaneously in order to obtain information about the
stimulus. For some specific responses, however, one of the aspects may be enough to identify the
stimulus. But on average, aspects cooperate. On the other hand, the presence of redundancy
($\Delta_{SR}<0$) indicates that, for many responses, the information conveyed by both aspects
overlaps. Therefore, some of the information that can be extracted from one aspect taken alone can
also be extracted from the other aspect taken alone. There might still be a few individual
responses for which it is necessary to read both aspects simultaneously to obtain information about
the stimulus. But on average, messages tend to be replicated in the different aspects.

In the absence of synergy/redundancy ($\Delta_{SR}=0$), the aspects might or might not be
independent and conditionally independent given the stimulus \citep{nirenberg2003, schneidman2003}.
If they are, then both aspects are fully autonomous. However, if they are not, then synergy and
redundancy coexist. Some responses might require the simultaneous read out of both aspects.
However, for other responses, at least one of the individual aspects might be enough to obtain
information about the stimulus. In this case, by considering both aspects separately, one cannot
recover the entire encoded information.

\subsection{Applications} \label{encdec}

The main ideas in this paper can also be extended to encompass any neuron response aspects,
different from pattern timing and pattern category. In particular, they allow us to analyse the
information conveyed by different types of patterns and the synergy/redundancy between them,
extending the formalism derived in \citet{eyherabide2008}. In addition, aspects may also be defined
in continuous time since \citep{mackay1952}. Even more, any neural population response can be
represented as a single sequence of coloured spikes, each colour indicating the neuron that fired
the spike \citep{brown2004}. Therefore, single neuron codes and population codes can be analysed
under the same formalism.

For example, during the last decades, many studies have focused on assessing whether different
neurons transmit information about different stimulus aspects \citep{gawne1996, denning2005,
eyherabide2008}. To that aim, different neurons (and different neural response aspects) have been
interpreted as information channels \citep{dan1998, montemurro2008, krieghoff2009}, often
addressing whether they constitute independent channels of information (see \ref{result:ife}).
However, these studies have focused on whether the two aspects (or neurons) are independent and
conditionally independent given the stimulus \citep{gawne1993, schneidman2003}. Indeed, in this
case, the response aspects constitute independent channels of information. However, these
conditions do not identify which stimulus aspects are encoded by different neurons, nor they
guarantee that they are independent.

To gain insight on the relation between stimulus and response aspects, we determine whether the
neuron constitutes a canonical feature extractor and/or a canonical feature interpreter. For independent
channels of information, the response aspects are canonical feature extractors, canonical feature interpreters,
and also independent and conditionally independent given the stimulus. However, none of these
conditions can be derived from the other. In effect, a canonical feature extractor or a canonical feature interpreter may or may not exhibit synergy or redundancy between the time and category information (see
\ref{result:ife} and \ref{appxH}). Moreover, even if $\mbf{T}$ and $\mbf{C}$ are independent and
conditionally independent given the stimulus, $\mbf{T}$ ($\mbf{C}$) may still convey information
about $\mbf{S_C}$ ($\mbf{S_T}$) once the information about $\mbf{S_T}$ ($\mbf{S_C}$) has been read
out. Formally, each of the equalities defining a canonical feature extractor or a canonical feature interpreter constitutes a relation between one aspect of the stimulus and one aspect of the response.
Such relations cannot be derived from the independence or conditional independence between two
aspects of the response. For the same reason, the \emph{what} and the \emph{when} are not
guaranteed to be independent aspects.

Finally, the analysis performed in this work relies on the mutual information between the stimulus
and different aspects of neural response, and thus it is related to both the encoding operation and
the decoding operation \citep{shannon1948, brown2004, nelken2007}. Indeed, the mutual information
is symmetric by definition \citep{coverthomas}. Therefore, one can interpret the information
between the stimulus and a particular aspect of the neural response from both points of view,
characterizing both how the stimulus is encoded into a specific response aspect \citep{reich2000,
reinagel2000, nelken2005, eyherabide2008, gollisch2008} and what can be inferred about the stimulus
from a decoder that only decodes that specific aspect. To this aim, an explicit representation of
each response aspect is needed, for example, as defined in \ref{methods:representations}. Such
representations are not always available, as for example, in studies assessing the role of
correlations (see \ref{difapp}).

\section*{Acknowldgements}

This work was supported by the Alexander von Humboldt Foundation, the Consejo de Investigaciones
Cient\'{\i}ficas y T\'ecnicas, the Agencia de Promoci\'on Cient\'{\i}fica y Tecnol\'ogica of
Argentina, and Comisi\'on Nacional de Energ\'{\i}a At\'omica.

\section*{Appendices}
\renewcommand{\thesubsection}{\Alph{subsection}.}
\setcounter{subsection}{0}
\renewcommand{\theequation}{\Alph{subsection}-\arabic{equation}}

\labelformat{subsection}{Appendix~\Alph{subsection}}

\subsection{Categorical noise} \label{appxA}
\setcounter{equation}{0}

The categorical noise is characterised by the probability $P_{\mbf{b}}(b|s)$ that a stimulus
feature $s$ elicits a pattern response of category $b$. This probability is related to
$P_{\mbf{r}}(\mbf{r}|s)$, the probability of inducing the response $\mbf{r}$ due to the feature
$s$, according to

\begin{equation}
P_{\mbf{b}}(b|s)=\sum_{\substack{\mbf{r}\\b=h_{\mbf{R}\rightarrow\mbf{B}}(\mbf{r})}}{
P_{\mbf{r}}(\mbf{r}|s)}\, ,
\end{equation}

\noindent where the sum runs through all patterns of spikes $\mbf{r}$ which category is $b$.

\subsection{Event counts transmit information at a vanishing rate} \label{appxB}
\setcounter{equation}{0}

Previous studies have shown that the information per unit time carried by the spike count decreases
with the size of the response time window \citep{petersen2002, montemurro2007b}. In this appendix,
we formally prove this result and also that the information per unit time vanishes in the limit of
long windows. We extend its validity not only for spikes, but for any response patterns, as defined
in \ref{methods:representations}, irrespective of the number of pattern categories. To that aim,
consider a representation $\eta$ that only preserves the number of patterns in each response
segment $\mbf{R}_{\tau}$ of length $\tau$ (Figure~\ref{f5}A). In this representation, two responses
stretches $\mbf{R}_{\tau}^1$ and $\mbf{R}_{\tau}^2$ are different if and only if they contain a
different number of patterns ($\eta(\mbf{R}_{\tau}^1) \neq \eta(\mbf{R}_{\tau}^2)$), otherwise they
are equal.

In a real experiment, patterns (and spikes) are not instant \citep{mackay1952}. Thus, without loss
of generality, consider the time divided into time bins of size $\Delta t$ shorter than the
shortest pattern. The number of events present in any response stretch $\mbf{R}_w$ of length $w$
bins is bounded by $0\leq\eta_w\leq w$, and therefore $\mathrm{H}(\eta_w)\leq \log{(w+1)}$. Hence,
the entropy rate $H(\eta)$ becomes zero, since

\begin{equation} \label{noentcount}
H(\eta)=\lim_{w\rightarrow\infty}{\frac{\mathrm{H}(\eta_w)}{w}}\leq
\lim_{w\rightarrow\infty}{\frac{\log{(w+1)}}{w}}=0\, .
\end{equation}

\noindent As a result, the information rate carried by $\eta$ about any other random variable
vanishes. In particular, $I(\mbf{S};\eta)\leq H(\eta) = 0$. The result is valid for response
patterns of any nature (see \ref{methods:representations} for the definition and examples of
patterns).

\subsection{The response set of event categories transmits information at a vanishing rate} \label{appxC}
\setcounter{equation}{0}

In this appendix, we prove that the information per unit time transmitted by the response set of
pattern categories decreases with the length of the response time window, and it vanishes in the
limit of long time windows. To this aim, we consider a representation $\Theta$ in which two
response segments are indistinguishable if and only if they have the same pattern categories,
irrespective of their temporal ordering  (see Figure \ref{f5}B). Hence, two neural responses can be
different in the category representation and equal in the $\Theta$ representation. Analogously to
\ref{appxB}, we only consider that the response events are not instant.

We first prove the result for the case where the number $|\Sigma_{\mbf{b}}|$ of possible different
pattern categories is finite; a neural response $\mbf{B}$ may be composed of several response
patterns. This is indeed the most frequent situation in the real neural system \citep{mackay1952},
valid for all the examples of pattern-based codes mentioned in \ref{methods:representations} and
throughout this paper. Consider that the neural response is read with words of length $w$ bins,
smaller than the shortest pattern. The number of patterns is bounded by $0\leq \eta_w \leq w$ (see
\ref{appxB}). In addition, each response pattern may belong to one out of $|\Sigma_{\mbf{b}}|$
pattern categories. Thus, the number of possible different responses $\Theta_w$ in the
representation $\Theta$ is upper-bounded by $|\Theta_w| \leq (w+1)^{|\Sigma_{\mbf{b}}|}$
\citep{coverthomas}. As a result, its entropy is upper-bounded by $\mathrm{H}(\Theta_w)\leq
\log{|\Theta_w|} $, and its entropy rate is

\begin{equation}
H(\Theta)=\lim_{w\rightarrow\infty}{\frac{\mathrm{H}(\Theta_w)}{w}}\leq
\lim_{w\rightarrow\infty}{|\Sigma_{\Theta}| \frac{\log{(w+1)}}{w}}=0\, .
\end{equation}

\noindent Therefore, there is no mutual information rate between the response set of pattern
categories and any other random variable. Particularly, $I(\mbf{S};\Theta)=0$.

We now generalise the result for infinite codes, under the only condition that patterns belonging
to different categories have different durations. These codes can be regarded as academic examples
since, in any real condition, they would be impractical due to the long time periods required to
read out the codewords. Examples of such infinite codes are bursts codes with no restriction in
their duration, inter-spike intervals or latencies divided into an infinite number of finite ranges
and the number of spikes in arbitrarily long time response windows. In a neural response of size
$w$ bins, only patterns up to a length $w$ can be read (see \ref{dpidiscuss} for examples). In
addition, a neural response may contain several patterns. Thus, the sum of the length of the
patterns cannot be greater than the length of the response containing them. Under this conditions,
the number of response sets of pattern categories $|\Theta_w|$ is upper-bounded by

\begin{equation}
|\Theta_w|\leq \sum_{k=0}^{w}{p(k)}\, ;
\end{equation}

\noindent where $p(k)$ represents the number of partitions of the integer number $k$. By using the
\emph{Hardy-Ramanujan-Uspensky} asymptotic approximation \citep{apostol1990}

\begin{equation}
|\Theta_w| \phantom{a} \leq \phantom{a} \sum_{k=0}^{w}{p(k)} \approx  C +
\sum_{k=k_0}^{w}{\frac{\mathrm{e}^{A\, \sqrt{k}}}{B\, k}} \phantom{a} \leq \phantom{a}
\frac{\mathrm{e}^{A\, \sqrt{w}}}{B}\, ;
\end{equation}

\noindent where $A$ and $B$ are positive constants, $k_0$ represents an integer for which the
approximation is valid, $C=\sum_{k=0}^{k=k_0 -1}{p(k)}$ and the right-most inequality is valid for
long enough words. Therefore, the entropy rate $H(\Theta)$ results

\begin{subequations}
\begin{align}
H(\Theta) &= \lim_{w\rightarrow\infty}{\frac{\log(|\Theta_w|)}{w}}\\
&\approx \lim_{w\rightarrow\infty}{\frac{A\, \log(\mathrm{e})}{\sqrt{w}}}-\lim_{w\rightarrow\infty}{\frac{\log(B)}{w}}\\
& = 0\, .
\end{align}
\end{subequations}

\noindent Thus, the entropy rate $H(\Theta)$ tends to zero, and consequently the mutual information
rate that the response set of pattern categories can carry about any other random variable
vanishes.

\subsection{Decomposition of the pattern information} \label{appxD}
\setcounter{equation}{0}

As mentioned previously, the pattern sequence $\mbf{B}$ and the pair $(\mbf{T}, \mbf{C})$ carry the
same information about the stimulus, since they are related through a bijective transformation.
Therefore

\begin{subequations}
\begin{align}
I(\mbf{S};\mbf{B}) &= I(\mbf{S};\mbf{T}, \mbf{C})\\
&= I(\mbf{S};\mbf{T}) + I(\mbf{S};\mbf{C}|\mbf{T})+I(\mbf{S};\mbf{C})-I(\mbf{S};\mbf{C})\label{infosplitap1}\\
&= I(\mbf{S};\mbf{T}) + I(\mbf{S};\mbf{C}) - (I(\mbf{S};\mbf{C})-I(\mbf{S};\mbf{C}|\mbf{T})) \label{infosplitap2} \\
&= I(\mbf{S};\mbf{T}) + I(\mbf{S};\mbf{C}) - I(\mbf{S};\mbf{T};\mbf{C}) \label{infosplitap3} \\
&= I(\mbf{S};\mbf{T}) + I(\mbf{S};\mbf{C}) + \Delta_{SR}\, ; \label{infosplitap4}
\end{align}
\end{subequations}

\noindent where Eq. \ref{infosplitap4} is obtained from Eq. \ref{infosplitap3} by replacing
$\Delta_{SR}= - I(\mbf{S};\mbf{T};\mbf{C})$. Here, $I(X;Y;Z)=I(X;Y)-I(X;Y|Z)$ represents the triple
mutual information \citep{coverthomas, tsujishita1995}.

\subsection{Relation between the upper- and lower-bounds of $\Delta_{SR}$}\label{appxE}

The synergy/redundancy term ($\Delta_{SR}$), defined in Eq. \ref{SynRed}, can be written as

\begin{subequations}\label{eqssinred}
\begin{align}
\Delta_{SR}&= I(\mbf{S};\mbf{T}|\mbf{C})-I(\mbf{S};\mbf{T})\label{SynRedTemp}\\
&= I(\mbf{S};\mbf{C}|\mbf{T})-I(\mbf{S};\mbf{C}) \label{SynRedCat}\\
&= I(\mbf{T};\mbf{C}|\mbf{S})-I(\mbf{T};\mbf{C})\, .\label{SynRedCatTemp}
\end{align}
\end{subequations}

\noindent Hence, the upper- and lower-bounds of $\Delta_{SR}$ are

\begin{equation}\label{synredboundsapp}
-\min\left\{\hspace{-0.15cm}\begin{array}{c}I(\mbf{T};\mbf{C})\\I(\mbf{S};\mbf{T})\\I(\mbf{S};\mbf{C})\end{array}\hspace{-0.15cm}\right\}
\leq \Delta_{SR} \leq
\min\left\{\hspace{-0.15cm}\begin{array}{c}I(\mbf{T};\mbf{C}|\mbf{S})\\I(\mbf{S};\mbf{T}|\mbf{C})\\I(\mbf{S};\mbf{C}|\mbf{T})\end{array}\hspace{-0.15cm}\right\}\,
.
\end{equation}

\noindent In addition, these upper- and lower-bounds are related through Eqs. \ref{eqssinred}, in
such a way that

\begin{equation}\label{synredboundsrel}
\min\left\{\hspace{-0.15cm}\begin{array}{c}I(\mbf{T};\mbf{C})\\I(\mbf{S};\mbf{T})\\I(\mbf{S};\mbf{C})\end{array}\hspace{-0.15cm}\right\}
= I(X, Y) \Leftrightarrow
\min\left\{\hspace{-0.15cm}\begin{array}{c}I(\mbf{T};\mbf{C}|\mbf{S})\\I(\mbf{S};\mbf{T}|\mbf{C})\\I(\mbf{S};\mbf{C}|\mbf{T})\end{array}\hspace{-0.15cm}\right\}=I(X,
Y|Z)\, ;
\end{equation}

\noindent where $X$, $Y$ and $Z$ represent the variables $\mbf{S}$, $\mbf{T}$ and $\mbf{C}$ in any
order. This proves the upper- and lower-bounds for the synergy/redundancy term $\Delta_{SR}$ of Eq.
\ref{synredbounds}.

\subsection{Information decomposition in representations of the neural response}\label{appxF}
\setcounter{equation}{0}

The information that a representation $\mbf{X}$ of the neural response conveys about the stimulus
can be decomposed as

\begin{equation}
I(\mbf{S};\mbf{X})=I(\mbf{S_T};\mbf{X})+I(\mbf{S_C};\mbf{X})+\Delta_{SR}^{X}\, ;
\end{equation}

\noindent where $I(\mbf{S_T};\mbf{X})$ is the information conveyed by $\mbf{X}$ about the
\emph{when} in the stimulus, and $I(\mbf{S_C};\mbf{X})$ is the information conveyed by $\mbf{X}$
the \emph{what}. Here, $\Delta_{SR}^X$ represents the synergy/redundancy between the information
conveyed about the \emph{when} and the \emph{what}, and it is given by

\begin{equation} \label{appxE1:delta}
\Delta_{SR}^X =-I(\mbf{S_T};\mbf{S_C};\mbf{X})\, ;
\end{equation}

\noindent which is lower-bounded by the redundancy in the stimulus

\begin{equation}
\Delta_{SR}^X \geq -I(\mbf{S_T};\mbf{S_C})\label{appxE1:delta2} \, .
\end{equation}

\noindent Tighter upper and lower bounds for $\Delta_{SR}^X$ can be derived analogously to the ones
derived for $\Delta_{SR}$ (see Eq. \ref{synredbounds}), as well as analogous conditions for the
absence of either synergy or redundancy between $I(\mbf{S_T};\mbf{X})$ and $I(\mbf{S_C};\mbf{X})$.

Notice that when $\Delta_{SR}^X>0$, the information provided about the stimulus is greater than the
sum of the information about \emph{when} and \emph{what} stimulus features happen, i.e.

\begin{equation}
I(\mbf{S};\mbf{X}) > I(\mbf{S_T};\mbf{X})+I(\mbf{S_C};\mbf{X})\, .
\end{equation}

\noindent This may occur, for example, if the latency in the response depends on the feature
category. In this case, the information that the time representation $\mbf{T}$ carries about the
time positions of stimulus features $\mbf{S_T}$ might be increased due to the knowledge of the
feature categories $\mbf{S_C}$. In conclusion, there would be an information component that is not
uniquely related to either \emph{when} or \emph{what}: It refers to both.

In the case that $I(\mbf{S_C};\mbf{X}|\mbf{S_T})=0$, the synergy/redundancy $\Delta_{SR}^X$ becomes

\begin{equation}
\Delta_{SR}^X =-I(\mbf{S_C};\mbf{X}) \label{appxE1:delta5} \, .
\end{equation}

\noindent Thus, $I(\mbf{S_C};\mbf{X})$ is completely redundant with and lower than
$I(\mbf{S_T};\mbf{X})$. That is,

\begin{subequations}\label{appxE1:delta6}
\begin{align}
I(\mbf{S};\mbf{X})&=I(\mbf{S_T};\mbf{X})+I(\mbf{S_C};\mbf{X})+\Delta_{SR}^{X}\\
I(\mbf{S_C};\mbf{X})+I(\mbf{S_T};\mbf{X}|\mbf{S_C})&=I(\mbf{S_T};\mbf{X})\\
I(\mbf{S_C};\mbf{X})&\leq I(\mbf{S_T};\mbf{X})\, .
\end{align}
\end{subequations}

Notice that this is the case of the time representation in a canonical feature extractor. Indeed, the
definition of a canonical feature extractor states that $I(\mbf{S_C};\mbf{T}|\mbf{S_T})=0$, and
consequently

\begin{equation}
\Delta_{SR}^T=-I(\mbf{S_C};\mbf{T})\, .
\end{equation}

\noindent The implications of condition \ref{infotempideal} mentioned in \ref{result:ife} follow
directly from this equation. By interchanging $\mbf{S_T}$ and $\mbf{S_C}$ in
Eqs.~\ref{appxE1:delta5} and \ref{appxE1:delta6}, analogous conclusions can be derived for the
category representation.

\subsection{Redundancy bounds for the canonical feature extractor} \label{appxG}
\setcounter{equation}{0}

To prove Eq. \ref{boundsynred}, we expand

\begin{subequations}
\begin{align}
&{}I(\mbf{T}, \mbf{S_T};\mbf{C}, \mbf{S_C}) = \nonumber\\
&=I(\mbf{S_T};\mbf{S_C})+\underbrace{I(\mbf{T};\mbf{S_C}|\mbf{S_T})}_{=0 \phantom{a}(Eq.
\ref{infotempideal})}+\underbrace{I(\mbf{S_T};\mbf{C}|\mbf{S_C})}_{=0 \phantom{a}(Eq.
\ref{infocatideal})}
+I(\mbf{T};\mbf{C}|\mbf{S_T}, \mbf{S_C})\label{appxE01}\\
&=I(\mbf{T};\mbf{C})+\underbrace{I(\mbf{S_T};\mbf{C}|\mbf{T})}_{\geq
0}+\underbrace{I(\mbf{T};\mbf{S_C}|\mbf{C})}_{\geq 0}+\underbrace{I(\mbf{S_T};\mbf{S_C}|\mbf{T},
\mbf{C})}_{\geq 0}\, .\label{appxE02}
\end{align}
\end{subequations}

Applying both conditions \ref{infotempideal} and \ref{infocatideal}, the second and third term of
Eq. \ref{appxE01} vanish respectively, and the synergy/redundancy between the time and category
information ($\Delta_{SR}$) is lower-bounded by

\begin{equation}
-I(\mbf{S_T};\mbf{S_C})\leq \Delta_{SR}\, .
\end{equation}

\noindent This is the bound that we wanted to prove.

\subsection{The canonical feature interpreter}\label{appxH}

We define a \emph{canonical feature interpreter} as a neuron model in which

\begin{subequations} \label{ifenc}
\begin{align}
I(\mbf{C};\mbf{S_T}|\mbf{T})&=0 \\
I(\mbf{T};\mbf{S_C}|\mbf{C})&=0 \, .
\end{align}
\end{subequations}

\noindent Under each of these conditions, the information conveyed by the neural response $\mbf{B}$
about the \emph{what} ($I(\mbf{B};\mbf{S_C})$) and about the \emph{when} ($I(\mbf{B};\mbf{S_T})$)
becomes

\begin{subequations}\label{ifencbounds}
\begin{align}
I(\mbf{B};\mbf{S_T})&=I(\mbf{T};\mbf{S_T})\leq H(\mbf{S_T})\\
I(\mbf{B};\mbf{S_C})&=I(\mbf{C};\mbf{S_C})\leq H(\mbf{S_C})\, .
\end{align}
\end{subequations}

\noindent Consequently, the \emph{what} (the \emph{when}) in the stimulus is completely represented
in the category (time) representation. In other words, $I(\mbf{S_C};\mbf{T})$
($I(\mbf{S_T};\mbf{C})$) is completely redundant with $I(\mbf{S_C};\mbf{C})$
($I(\mbf{S_T};\mbf{T})$), and $I(\mbf{S_C};\mbf{T}) \leq I(\mbf{S_C};\mbf{C})$
($I(\mbf{S_T};\mbf{C}) \leq I(\mbf{S_T};\mbf{T})$). The canonical feature interpreter is analogous to the
canonical feature extractor. In fact, it can be obtained by interchanging the role of the stimulus and
the response in \ref{result:ife}.

\footnotesize
\bibliography{EyherabideSamengo-References}

\end{document}